\documentclass[a4paper]{svjour3}                     
\smartqed  
\usepackage{amssymb}
\usepackage[pdftex]{graphicx}
\usepackage[labelsep=space,labelfont=bf]{caption}
\usepackage{enumerate}
\usepackage{color}
\usepackage{amsmath}

\newcommand{\fk}[1]{\textit{\textbf{#1}}}
\newcommand{\figref}[1]{Fig. \ref{#1}}
\newcommand{\secref}[1]{Sect. \ref{#1}}
\newcommand{\figtext}[1]{\textit{\small{#1}}}

\makeatletter
\newcommand*{\rom}[1]{\expandafter\@slowromancap\romannumeral #1@}
\makeatother

\journalname{}
\begin{document}

\title{A Hydrodynamic Model of Movement of a Contact Line Over a Curved Wall \thanks{The authors were supported by the Swedish Research Council.}
}


\author{Hanna Holmgren         \and
       Gunilla Kreiss
}


\institute{Hanna Holmgren \at
              Department of Information Technology, Uppsala University, Box 337, SE-751 05, Uppsala, Sweden \\
              Tel.: +46-761287654\\
              \email{hanna.holmgren@it.uu.se}           
           \and
           Gunilla Kreiss\at
              Department of Information Technology, Uppsala University, Box 337, SE-751 05, Uppsala, Sweden 
}

\date{}


\maketitle

\begin{abstract}
The conventional no-slip boundary condition leads to a non-integrable stress singularity at a contact line. This is a main challenge in numerical simulations of two-phase flows with moving contact lines. We derive a two-dimensional hydrodynamic model for the velocity field at a contact point moving with constant velocity over a curved wall. The model is a perturbation of the classical Huh and Scriven hydrodynamic solution \cite{HUH}, which is only valid for flow over a flat wall. The purpose of the hydrodynamic model is to investigate the macroscopic behavior of the fluids close to a contact point. 
We also present an idea for how the hydrodynamic solution could be used to prescribe macroscopic Dirichlet boundary conditions for the velocity in the vicinity of a moving contact point. Simulations demonstrate that the velocity field based on the non-singular boundary conditions is capable of accurately advecting the contact point. 
\end{abstract}

\newpage
\section{Introduction}
Consider two immiscible and incompressible  fluids. If the fluids are in contact with a solid there is a line where the interface separating the two fluids meet the solid. This line is called a contact line, and when it evolves with time we have a  moving contact line problem. Such problems form an important class of two-phase flows and appear both in nature and in many industrial applications \cite{REW}. Examples of phenomena in nature are raindrops falling on a window or water bugs resting on water surfaces \cite{LIU}. Industrial applications include lubrication, inkjet printing, gas and oil recovery in porous media \cite{ZAHEDI,ROCCA,1ROCCA,2ROCCA,3ROCCA,4ROCCA} and the development of microfluidic devices such as micropumps and lab-on-a-chip devices \cite{MARTIN}. 

The standard mathematical model is the incompressible Navier-Stokes equations influenced by the interface through surface tension. The interface in its turn is advected by the fluid velocity. A main challenge in numerical simulations of moving contact line problems is that the adherence, or no-slip boundary condition leads to a non-integrable stress singularity at the contact line \cite{REW,DUSSAN1}. With a no-slip condition an infinite force would be required to move the contact line. However, the surface tension force, which is believed to cause the contact line movement, is finite \cite{RENARDY}. 
To accurately model contact line dynamics atomistic phenomena must be taken into account in a small region near the contact line. 

In their paper Huh and Scriven \cite{HUH} argue that even though the ultimate resolution of a moving contact line must rest on molecular considerations on a microscopic length scale, the problem can be approached through kinematics and dynamics of fluids. They mean that it is still instructive to use a hydrodynamical model to investigate the macroscopic behavior of the fluids close to, but not exactly at, a contact line.
In \cite{HUH} Huh and Scriven construct such a hydrodynamical model for the case of steady, two-dimensional flow with a perfectly flat fluid interface. The model is aimed at describing the flow at an intermediate length scale $L$, which is much smaller than the large scale features of the flow but still large enough for the atomistic phenomena to be negligible. 
The model is based on the assumption that at this intermediate scale the viscous effects dominates the convection and the creeping flow approximation of Navier--Stokes equations is valid. The analysis results in a family of solutions, which depends on three parameters: the velocity of the contact point (in 2D the contact line is reduced to a contact point), the contact angle (the angle between the interface and the wall), and the viscosity ratio of the two fluids.


The Huh and Scriven  solution is referred to extensively in the fluid dynamics community, and it is recognized to be a useful tool for describing flow in an intermediate region near contact points, see for example the recent review paper by Snoijer and Andreotti \cite{Review2013}, and also \cite{Yulii,DUSSAN1,DUSSAN}. It is also well recognized that this solution has shortcomings. Firstly, the Huh and Scriven  solution is singular at the contact point. It is by now well known that at the contact point the Navier--Stokes equations do not model the physics. In this small region atomistic phenomena come into play, and these are not included in the standard Navier--Stokes system. Even if the Huh and Scriven  solution were regular there, it would not be physically relevant. The Huh and Scriven  solution is relevant at the intermediate length scale, much smaller than that of main features in the flow, but larger than the atomistic scale. Secondly, the assumption of a planar interface is unrealistic. There is a jump in the pressure over the interface, which can only be balanced by surface tension of a curved surface. However, if the surface tension effect is strong, which is the case for capillary dominated flow, a small curvature suffices. In \cite{chen1997} a modified Huh and Scriven solution is presented, which reveals that at an intermediate length scale the curvature decreases away from the contact point, and the flow approaches the flow of the planar case. There is also experimental agreement. Further away from the contact point, the assumption of a flat fluid interface, as well as the assumption of a low Reynolds number, are in general no longer valid. The conclusion is that the planar interface and the Huh and Scriven  solution are appropriate for matching an outer solution with an inner solution, at some intermediate distance from the contact point. 
Thus, important to note is that the Huh and Scriven  solution on its own does not suffice for determining the movement of the contact point, it must be combined with a model taking the atomistic processes into account. Examples of such models are the phenomenologically based phase field model, or a molecular dynamics
based model. 

The main focus of this work is the modification of the hydrodynamical model derived by Huh and Scriven to be valid for a different geometry. In \cite{HUH} the solution is derived for a contact point moving over a flat wall. A flat wall is often considered in numerical simulations of moving contact line problems because of its simplicity \cite{LIU}. However, in many applications more complex geometries need to be taken into account \cite{LIU}. 


In the first part of this paper, we extend the hydrodynamical model presented in \cite{HUH} to a circular solid boundary with radius of curvature $R$. 
If the curvature of a more complex boundary varies sufficiently slow, the boundary can be well approximated to have a circular shape locally at the contact point, and our model can be used.
Further, by assuming $R>L$ we can neglect centrifugal and Coriolis forces.
Since we are interested in the dynamics close to the contact point we use perturbation analysis over the small quantity $r/R$, where $r<L$ is the distance to the contact point. In the first part of this paper we determine the first two terms in the expansion. However, it is possible to extend the theory and calculations presented here to account for more terms in the expansions if required.

In the second part of this paper we show one possible example of how our hydrodynamical model could be used in numerical simulations of moving contact point problems. The focus is on how to incorporate small-scale effects on the macro-dynamics. The small-scale effects are assumed to be given by a relation between the macroscale wall contact angle and the contact point speed. We present numerical experiments designed to test how well a velocity field with boundary conditions based on the extended hydrodynamic solution advects the contact point.

\section{PART \rom{1}: Derivation of the Hydrodynamic Model} 
\label{sec:PART1}
In this section we will construct a hydrodynamical model for the velocity field close to a contact point moving over a curved wall. In \secref{sec:hyd_model} we start by describing our two-dimensional model problem, and discuss the conditions under which the creeping flow approximation (also called the Stokes model) can be used in the vicinity of the contact point. In \secref{sec:perturbation} we introduce a perturbation expansion, and solve for the first two terms. Finally, \secref{sec:result1} discusses resulting velocity fields.

\subsection{The Contact Point Model Problem}
\label{sec:hyd_model}
\begin{figure}[h!]
  \centering
    \includegraphics[width=0.5\textwidth]{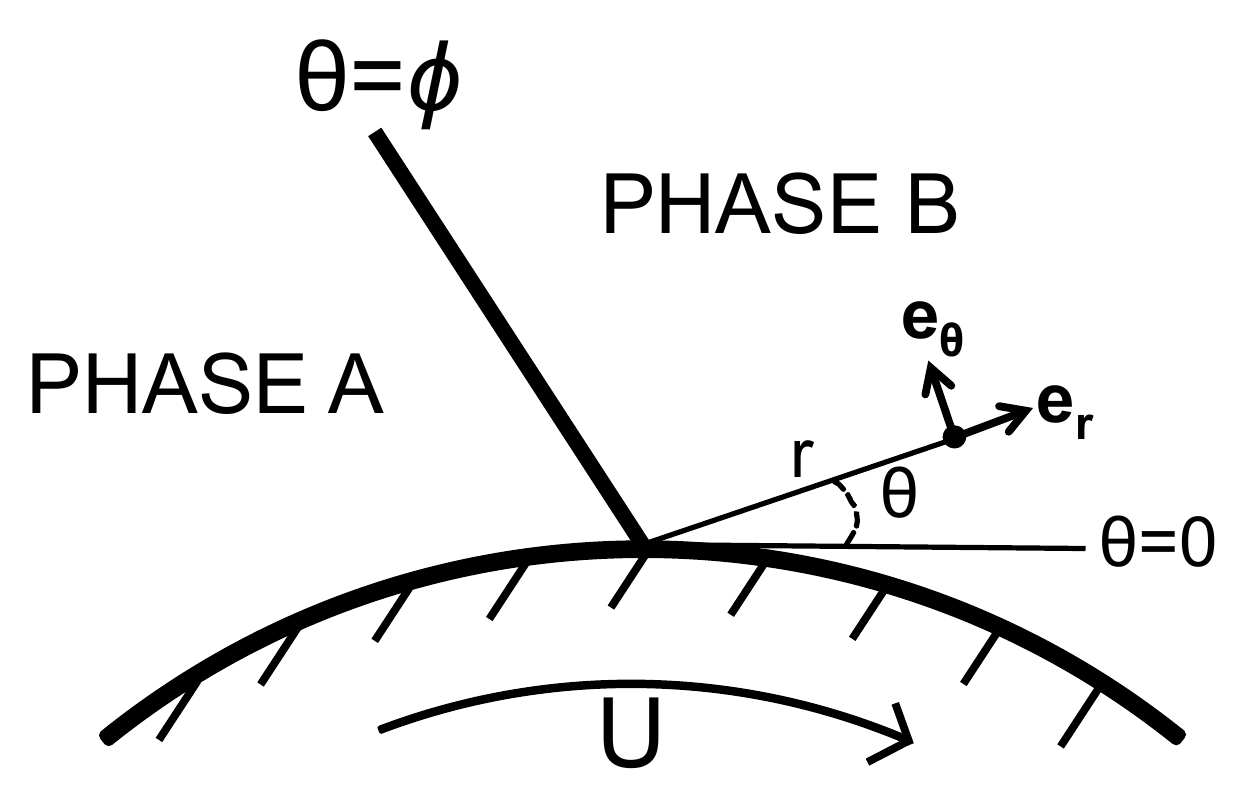}
    \caption{Model problem}
    \label{fig:setup}
\end{figure}

Consider a model problem in two space dimensions with constant curvature of the solid wall and constant contact point speed $U$, see \figref{fig:setup}. A polar coordinate system $(r, \theta)$ is used, with the origin  fixed to the contact point position, and with $\theta=0,\,180$  along the tangent of the wall at the contact point. This coordinate system is rotating with  angular velocity  ${\bf \Omega}$ of magnitude  $U/R$, where $R$ is the radius of curvature of the wall. In this rotating frame of reference the contact point appears to be at rest and the wall to move with constant speed $U$ (but with opposite direction to that of the contact point in a fixed frame). The fluid interface between phase A and B is assumed to be perfectly flat and inclined at an angle of $\phi$ from the line $\theta=0$. 

\subsubsection{Creeping flow approximation}
In a rotating frame of reference  centrifugal and Coriolis forces appear in the Navier-Stokes equations:
\begin{equation}\nonumber
\frac{\partial \fk{v}'}{\partial t'}+(\fk{v}'\cdot\nabla) \fk{v}'+\frac{1}{\rho}\nabla p'=\frac{\mu}{\rho}\nabla^2 \fk{v}'-2 {\bf \Omega}' \times \fk{v}' -{\bf \Omega} '\times({\bf \Omega}'\times r'\fk{e}_r).
\end{equation}
Here $\fk{v}'$ is the fluid velocity, $p'$ is the pressure, $\mu$ is the viscosity, and $\rho$ is the density.  Introduce dimensionless quantities by
\begin{equation} \fk{v}=U\fk{v}',\quad r=Lr' , \quad t=\frac{L}{U}t', \quad p= P p', \quad{\bf \Omega}=\frac{U}{R} {\bf \Omega}'. \end{equation}
Now the equations are
\begin{equation}\nonumber
\frac{U^2}{L}\frac{\partial \fk{v}}{\partial t}+\frac{U^2}{L}(\fk{u}\cdot\nabla) \fk{v}+\frac{P}{L\rho}\nabla p=\frac{\mu U}{\rho L^2}\nabla^2 \fk{v}-2\frac{U^2}{R}{\bf \Omega}\times \fk{v}-\frac{LU^2}{R^2}{\bf \Omega}\times({\bf \Omega}\times r\,\fk{e}_r)  .
\end{equation}
With the Reynolds number $\mathrm{Re}=LU\rho/\mu$ we have (after multiplying with $\frac{\rho L^2}{\mu U}$)
\begin{equation}\nonumber
\mathrm{Re}\frac{\partial \fk{v}}{\partial t}+\mathrm{Re}(\fk{v}\cdot\nabla) \fk{v}+\frac{PL}{\mu U}\nabla p=\nabla^2 \fk{v}-2\mathrm{Re}\frac{L}{R}{\bf \Omega} \times \fk{v}-\mathrm{Re} \frac{L^2}{R^2}{\bf \Omega}\times({\bf \Omega}\times r\,\fk{e}_r). 
\end{equation}

Close to the contact point viscous effects dominate over convection, and $\mathrm{Re}\ll 1$. Thus  we neglect inertia terms (first two terms above). In addition,
since $R> L$ we similarly  neglect the centrifugal and Coriolis forces (last two terms). The result is that we can use the same creeping flow equations as in  the flat wall case.

The creeping flow equations can be formulated to take the form of a two-dimensional biharmonic equation for the stream function $\psi(r,\theta)$ \cite{BIHARMONIC}:
\begin{align}\label{eq:biharmonic}
\nabla^4\psi=0.
\end{align}
In terms of the stream function the fluid velocity components in plane polar coordinates are:
\begin{align}
	&v_r=-\frac{1}{r} \frac{\partial \psi}{\partial \theta} \label{eq:u} \\
	&v_{\theta}= \frac{\partial \psi}{\partial r}.    \label{eq:v}
\end{align}
In our setting we will have one stream function in each fluid, $\psi_A$ and $\psi_B$, and corresponding velocities.

\subsubsection{Boundary and Interface Conditions}
To close the system \eqref{eq:biharmonic} we need boundary and interface conditions.
Before applying the boundary conditions, we parametrize the circular wall as:\small
\begin{equation}
\left.
\begin{aligned}
	\theta&= \arcsin (\textstyle\frac{r}{2R})+\pi ~~~\text{in A}  ~\\
	\theta&= -\arcsin (\textstyle\frac{r}{2R}) ~~~\text{in B} 
\end{aligned}
\right\}.
\label{eq:parametri}
\end{equation}
When applying the boundary conditions, the following reformulations of the parametrization in \eqref{eq:parametri} will also be needed: 
\begin{equation*}
	\sin \theta=-\textstyle\frac{r}{2R} \text{,  }~  \cos2\theta= 1-\textstyle\frac{r^2}{4R^2}   ~~~\text{in both A and B} 
\end{equation*}
\begin{equation}
\left.
\begin{aligned}
	\cos\theta&= -\textstyle\sqrt{1-\textstyle\frac{r^2}{4R^2}} ~~~\text{in A} ~ \\
	\cos\theta&=\textstyle\sqrt{1-\textstyle\frac{r^2}{4R^2}} ~~~\text{in B}
\end{aligned}
\right\}
\quad \quad ~
\left.
\begin{aligned}
	\sin2\theta&= \textstyle\frac{r}{R}\textstyle \sqrt{1-\frac{r^2}{4R^2}}  ~~~\text{in A} ~ \\
	\sin2\theta&= -\textstyle\frac{r}{R}\textstyle \sqrt{1-\frac{r^2}{4R^2}}   ~~~\text{in B}
\end{aligned}
\right\}  .
\label{eq:circ}
\end{equation}
\normalsize
Further, the normals and tangents to the circle are described in plane polar coordinates by \small
\begin{equation}
	\fk{n}=(-\sin \theta, \cos \theta) ~~\text{in both A and B} \quad \quad ~
	\left.
\begin{aligned}
	&\fk{t}_A=(-\cos \theta,-\sin \theta ) ~~~\text{in A} ~ \\
	&\fk{t}_B=(\cos \theta,\sin \theta ) ~~~\text{in B} \label{eq:tangent}
\end{aligned}
\right\}.
\end{equation}
\normalsize
Note that  the components of these vectors are in the radial and angular directions, respectively.

Now, the following boundary and interface conditions can be applied:
\begin{enumerate}[(i)]
	\item Kinematic conditions: vanishing normal component of velocity at the solid wall and fluid interface.
	\begin{align}
		& \text{\small At the circular wall given by \eqref{eq:parametri}:} \notag \\
		&\fk{n}\cdot \fk{v}_A = \sin\theta \frac{1}{r} \frac{\partial \psi_A}	{\partial \theta} + \cos\theta \frac{\partial \psi_A}	{\partial r}= 0  				\label{eq:BC1} \\
		\notag \\
		&\text{\small At the circular wall given by \eqref{eq:parametri}:} \notag \\
		&\fk{n}\cdot \fk{v}_B = \sin\theta \frac{1}{r} \frac{\partial \psi_B}	{\partial \theta} + \cos\theta \frac{\partial \psi_B}	{\partial r}= 0
		\label{eq:BC2}   \\
		\notag \\
		&\text{\small At the interface, i.e. $\theta=\phi$:}~~~ v_{A\theta} = \frac{\partial \psi_A}{\partial r} = 0 
		\label{eq:BC3}  \\
		\notag \\
		&\text{\small At the interface, i.e. $\theta=\phi$:}~~~ v_{B\theta} = \frac{\partial \psi_B}{\partial r} = 0 
		\label{eq:BC4}  \\
		\notag 
	\end{align}
	
	\item Kinematic condition: no slip, i.e. continuity of velocity at the interface.
	\begin{align}
		&\text{\small At the interface, i.e. $\theta=\phi$ ($r>0$):}~~~ v_{Ar}=v_{Br}  ~\Leftrightarrow   \notag \\
		&\frac{\partial \psi_A}{\partial \theta}=	\frac{\partial \psi_B}{\partial \theta} 
		\label{eq:BC5} 
	\end{align}
	
	\item Dynamic condition: continuity of tangential stress at the interface.
	\begin{align}
		& \text{\small At the interface, i.e. $\theta=\phi$ ($r>0$):}~~~ \tau_{Ar\theta}=\tau_{Br\theta}  ~\Leftrightarrow  \notag \\
		&~ \mu_A\frac{\partial^2 \psi_A}{\partial \theta^2}=\mu_B\frac{\partial^2 \psi_B}{\partial \theta^2}
		\label{eq:BC6} 
	\end{align}
	\small Equations \eqref{eq:BC3} and \eqref{eq:BC4} have been used to simplify the second expression.
		
	\item Kinematic condition: no slip, i.e., no tangential relative motion of the fluid at the solid wall.
	\begin{align}
		&\text{\small At the circular wall given by \eqref{eq:parametri}:} \notag \\
		&\fk{t}_A\cdot \fk{v}_A = \cos\theta \frac{1}{r} \frac{\partial \psi_A}	{\partial \theta} - \sin\theta \frac{\partial \psi_A}	{\partial r}= U.
		\label{eq:BC7}   \\
		\notag \\
		&\text{\small At the circular wall given by \eqref{eq:parametri}:} \notag \\
		&\fk{t}_B\cdot \fk{v}_B = -\cos\theta \frac{1}{r} \frac{\partial \psi_B}	{\partial \theta} - \cos\theta \frac{\partial \psi_B}	{\partial r}= -U  .				\label{eq:BC8}\\
		\notag 
	\end{align}
	
\end{enumerate}

\subsection{Approximate Solution by Perturbation Analysis}
\label{sec:perturbation}

In this section we use perturbation analysis to find an approximate expression for the velocity field close to a contact point at rest at  a circular wall, which rotates with constant speed.
We are looking for an expansion of the velocity field in terms  of different powers of  $r/R$,
\begin{equation}\label{eq:expansion1}
\fk{v}=v^0+\frac{r}{R}v^1+\dots .
\end{equation}

The starting point is the biharmonic equation (\ref{eq:biharmonic}) in plane polar coordinates for each fluid, together with the boundary and interface conditions \eqref{eq:BC1} - \eqref{eq:BC8}. The velocity  is given by derivatives of the stream function according to \eqref{eq:u} and \eqref{eq:v}.
The general solution of the biharmonic equation in plane polar coordinates was derived in a paper by Michell \cite{MICHELL} by a separation of variables. The solution is in the form of an expansion containing terms with different powers of $r$. Here,  just as in \cite{HUH} we require that the velocity is bounded as $r \rightarrow 0$, and therefore all terms  with negative powers of $r$ in the Michell solution are excluded. In \cite{HUH} the expansion was to be used also far from the contact point, and therefore the velocity was required to be bounded also for large values of $r$. The consequence was that all terms  with powers of $r$ higher than one in the expression for the stream function were also omitted. 
In our setting the expansion is only used close to the contact point. Further away it needs to be matched to an outer solution, and therefore there is no reason to omit terms with higher powers of $r$.  

Based on \cite{MICHELL} we use the following expansion of the stream function 
\begin{equation}
\begin{aligned}
	\psi (r,\theta)&=r(a \sin\theta+b\cos\theta+c\theta\sin\theta+d\theta\cos\theta ) +\\
	&+ \frac{r^2}{R}(e+f\cos2\theta+g\sin2\theta+h\theta)+\dots,
	\end{aligned}
	\label{eq:psi}
\end{equation}
for each of the fluids A and B.
For the stream function expansion to correspond to (\ref{eq:expansion1}), we have scaled the constants $e, f, g$ and $h$ in the second term above by $1/R$, compared to the solution given in \cite{MICHELL} (i.e. the constants $e, f, g$ and $h$ here equals the constants from the Michell solution multiplied by $R$). 
Further, in the paper by Michell the term $r^2\theta$ was not included due to an assumption of theta-periodicity, which is not assumed here. 

Using the expansion for the stream function (17), the zeroth and first order components (in $r/R$) of the expansion for the velocity (16) are given by
\begin{align}
&v^0_r=(b+d\theta -c)\sin\theta-(a+c\theta +d)\cos\theta   \label{eq:velcomp1} \\
&v^0_{\theta}=(a+c\theta)\sin\theta+(b+d\theta)\cos\theta  \label{eq:velcomp2} \\
&v^1_r=2f\sin2\theta-2g\cos2\theta-h  \label{eq:velcomp3} \\
&v^1_{\theta}=2(e+f\cos2\theta+g\sin2\theta+h\theta). \label{eq:velcomp4}
\end{align}
 for each of the fluids A and B. We will in the following determine the coefficients in the two leading order terms of the expansions for each of the fluids by applying the boundary conditions  \eqref{eq:BC1} - \eqref{eq:BC8}. Note that  it would be straight forward to continue the expansion to higher powers of $r/R$. 

To consider the boundary conditions along the circular wall to different powers of $r/R$, we need to taylor expand the parametrization of the circular wall given in \eqref{eq:parametri} and \eqref{eq:circ} around $r/R=0$:
\small 
\begin{equation*}
\left.
\begin{aligned}
	\theta&= \pi + \textstyle \frac{r}{2R} +  \mathcal{O}(\frac{r^3}{R^3}) ~~~\text{in A}  ~\\
	\theta&=  0   - \textstyle \frac{r}{2R} + \mathcal{O}(\frac{r^3}{R^3})     ~~~\text{in B} 
\end{aligned}
\right\}
\end{equation*}
\begin{equation*}
	\sin \theta=-\textstyle \frac{r}{2R} \text{, }~ \cos2\theta= 1 + 0 + \mathcal{O}(\frac{r^2}{R^2}) ~~~\text{in both A and B}  
\end{equation*}
\begin{equation}
\left.
\begin{aligned}
	\cos\theta&= -1 + 0 + \mathcal{O}( \textstyle\frac{r^2}{R^2})  ~~~\text{in A} ~ \\
	\cos\theta&=1 + 0 + \mathcal{O}( \textstyle\frac{r^2}{R^2})  ~~~\text{in B}
\end{aligned}
\right\}
\quad \quad 
\left.
\begin{aligned}
	\sin2\theta&= 0+ \textstyle\frac{r}{R} + \mathcal{O}(\frac{r^3}{R^3}) ~~~\text{in A} ~ \\
	\sin2\theta&= 0- \textstyle\frac{r}{R} + \mathcal{O}(\frac{r^3}{R^3})  ~~~\text{in B}
\end{aligned}
\right\}.
\end{equation}
 
 \normalsize
 
We are now ready to apply the boundary and interface conditions \eqref{eq:BC1} - \eqref{eq:BC8} to the expression for the stream function \eqref{eq:psi}, and consider separatly   different powers of $r/R$.
The zeroth order equations involve $[a_i, b_i, c_i, d_i]$, $i=A; B$, and  are  exactly the same as in \cite{HUH}. 
 It follows that  the velocity field derived in \cite{HUH} for a contact point at rest at a flat wall, which translates with constant velocity, is the unperturbed solution $v_0$. 
For clarity we have stated the equations for the coefficients $[a_i, b_i, c_i, d_i]$, $i=A; B$,  and their solutions,  in the appendix. We note that the system is singular for $\phi=0,180$, and non-singular for other angles.
 
Next we collect terms that are linear in $r/R$ and get the following system for the coefficients $[e_i, f_i, g_i, h_i]$, $i=A; B$ from the second term of (\ref{eq:psi}):
\begin{equation}
M^1Z^1=F^1.
\label{eq:first_syst}
\end{equation}
Here
\begin{equation}
M^1= 
\small
\begin{pmatrix}
 1 & 1 & 0 & \pi & 0 & 0 & 0 & 0  \\  
 0 & 0 & 0 & 0 & 1 & 1 & 0 & 0  \\    
 1 & \tilde{C} & \tilde{S} & \phi & 0 & 0 & 0 & 0  \\  
 0 & 0 & 0 & 0 & 1 & \tilde{C} & \tilde{S} & \phi  \\ 
 0 & 2\tilde{S} & -2\tilde{C} & -1 & 0 & -2\tilde{S} & 2\tilde{C} & 1  \\ 
 0 & \mu_A \tilde{C} & \mu_A\tilde{S} & 0 & 0 & -\mu_B\tilde{C} & -\mu_B\tilde{S} &0  \\  
 0 & 0 & -1 & -1/2 & 0 & 0 & 0 & 0 \\   
 0 & 0 & 0 & 0 & 0 & 0 & 1 & 1/2    
 \end{pmatrix} \notag, 
 \end{equation} 
 
 \normalsize
 \begin{equation}
 Z^1=
  \begin{pmatrix}		      
    e_A\\
    f_A  \\
    g_A  \\
    h_A  \\
    e_B \\
    f_B \\
    g_B \\
    h_B
 \end{pmatrix}
~~~\text{       and       } ~~~
F^1=
 \begin{pmatrix}		      
     \frac{1}{2}(a_A+\pi c_A+d_A)\\ 
     \frac{1}{2}(a_B+d_B) \\ 
    0  \\ 
    0  \\ 
    0 \\  
    0 \\  
     \frac{1}{2}(b_A-c_A +\pi d_A) \\  
     \frac{1}{2}(c_B - b_B) 
 \end{pmatrix}.
  \notag
\end{equation}
We have used the notations 
\begin{align}
&\tilde{S}=\sin2\phi \notag \\
&\tilde{C}=\cos2\phi .\notag 
\end{align} 
The rows in system \eqref{eq:first_syst} originate from the boundary and interface conditions \eqref{eq:BC1}-\eqref{eq:BC8}, respectively.
This system is also singular for precisely $\phi=0, 180$.

When all coefficients $[a_i, b_i, c_i, d_i, e_i, f_i, g_i, h_i]$, $i=A; B$ have been calculated the resulting approximated velocity field is obtained using the expressions \eqref{eq:velcomp1} - \eqref{eq:velcomp4}.

\subsection{Resulting Velocity Fields and Verification}
\label{sec:result1}
\begin{figure}[h!]
  \centering
    \hspace{-1.5cm} \includegraphics[width=0.8\textwidth]{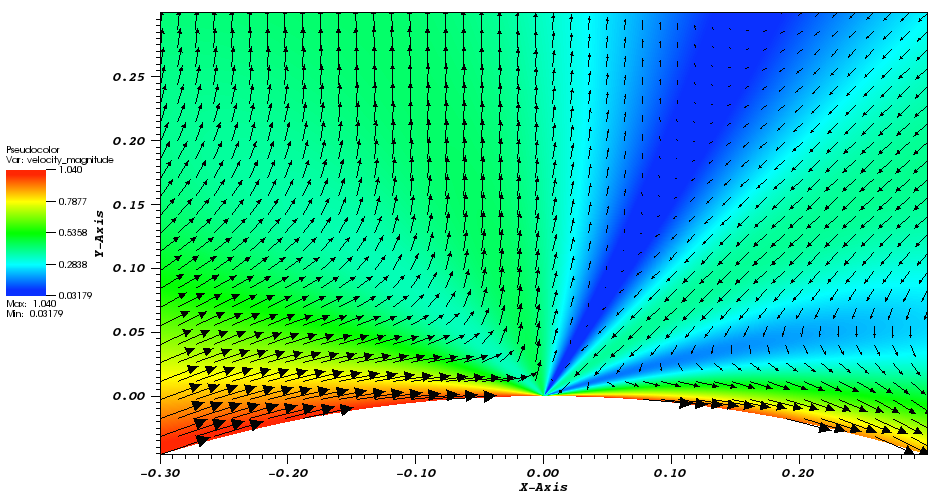}
    \caption{Approximated velocity field for the first case: $R=1$, $\mu_A/\mu_B=1$, $\phi=45$, $U=1$}
    \label{fig:vel_pert}
\end{figure}

\begin{figure}[h!]
  \centering
   \hspace{-1.5cm}  \includegraphics[width=0.8\textwidth]{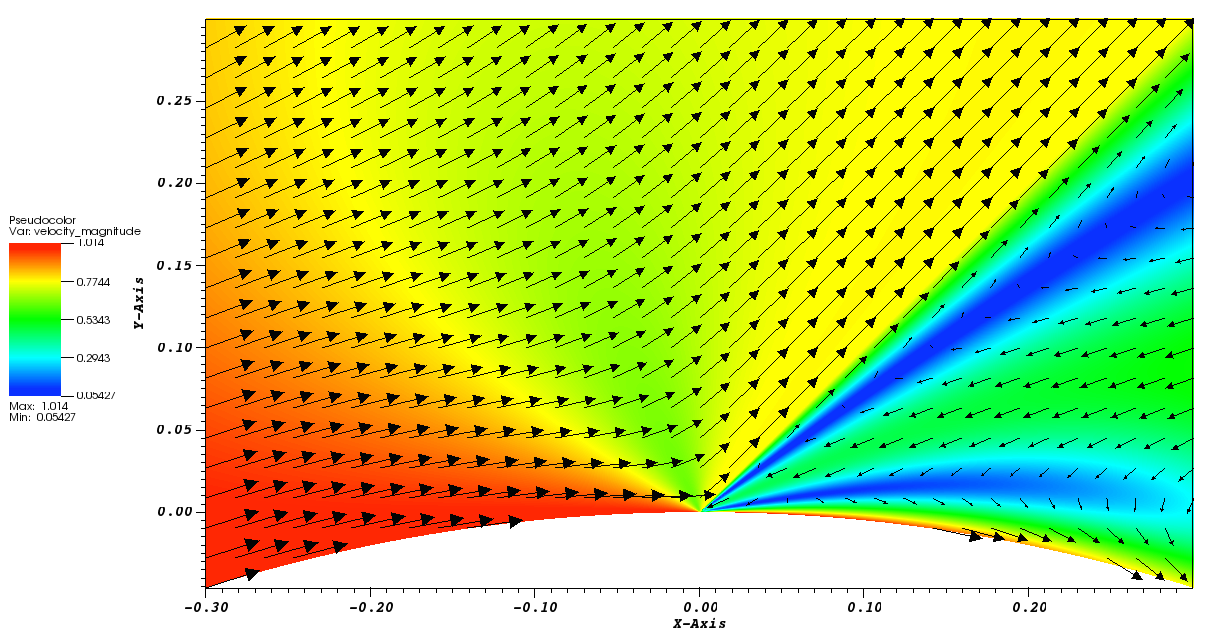}
    \caption{Approximated velocity field for the second case: $R=1$, $\mu_A/\mu_B=100$, $\phi=45$, $U=1$}
    \label{fig:vel_pert2}
\end{figure}

\begin{figure}[h!]
  \centering
    \hspace{-1.5cm} \includegraphics[width=0.8\textwidth]{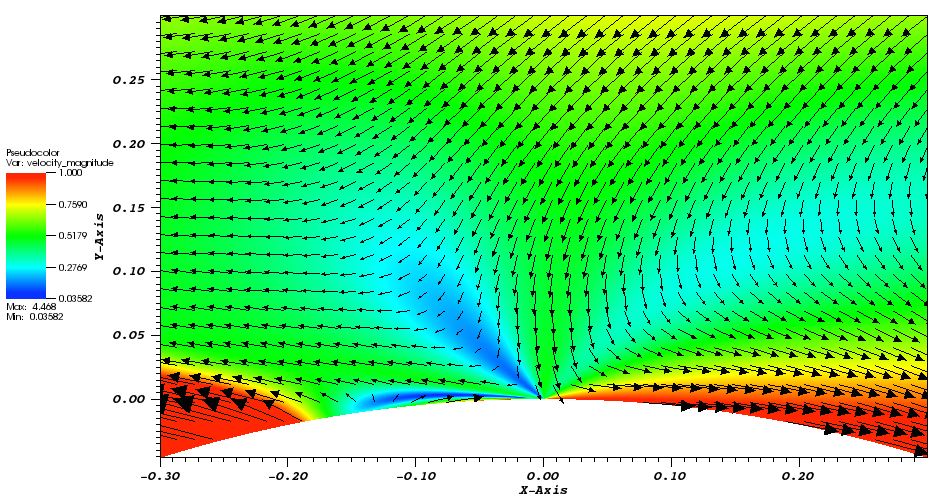}
    \caption{Approximated velocity field for the third case: $R=1$, $\mu_A/\mu_B=100$, $\phi=170$, $U=1$}
    \label{fig:vel_pert3}
\end{figure}

We look at the resulting approximated velocity field for three different cases. In all cases the contact point is 
 stationary while the circular solid of radius $R=1$ is rotating with angular velocity $U/R$ where $U=1$.  In the first case the angle between the fluid interface and the line $\theta=0$ is $\phi=45$ and the viscosity ratio is $\mu_A/\mu_B=1$. In the second case the viscosity ratio is increased to $\mu_A/\mu_B=100$. In the third case the higher viscosity ratio is used again, but the contact angle is now increased to $\phi=170$. The resulting velocity fields for the three cases are shown in  \figref{fig:vel_pert}, \figref{fig:vel_pert2} and  \figref{fig:vel_pert3} respectively.

Increasing the viscosity ratio from $\mu_A/\mu_B=1$ to $\mu_A/\mu_B=100$ leads to a significant change in the flow pattern, compare case one in \figref{fig:vel_pert} and case two in \figref{fig:vel_pert2}. In the first case, when the two phases have the same viscosity, the flow along the circular boundary in phase A and along the interface both are directed inwards towards the contact point. This leads to the formation of a jet out from the contact point in phase A (directed upwards in \figref{fig:vel_pert}). In the second case, when the viscosity ratio is increased to $\mu_A/\mu_B=100$, the direction of the flow along the interface has changed sign compared to the equal viscosity case. Consequently the flow along the circular boundary in phase B and along the interface are now both directed outwards from the contact point, which leads to a jet  inwards towards the contact point in phase B (the region with lower viscosity) instead. In case three, when the contact angle is increased to $\phi=170$ but the viscosity ratio is kept at $\mu_A/\mu_B=100$, the same flow pattern with an ingoing jet towards the contact point in the phase with lower viscosity, phase B, is observed. This behavior agrees with the observations in \cite{HUH}.


\begin{figure}[h!]
\centering
     \includegraphics[width=0.9\textwidth]{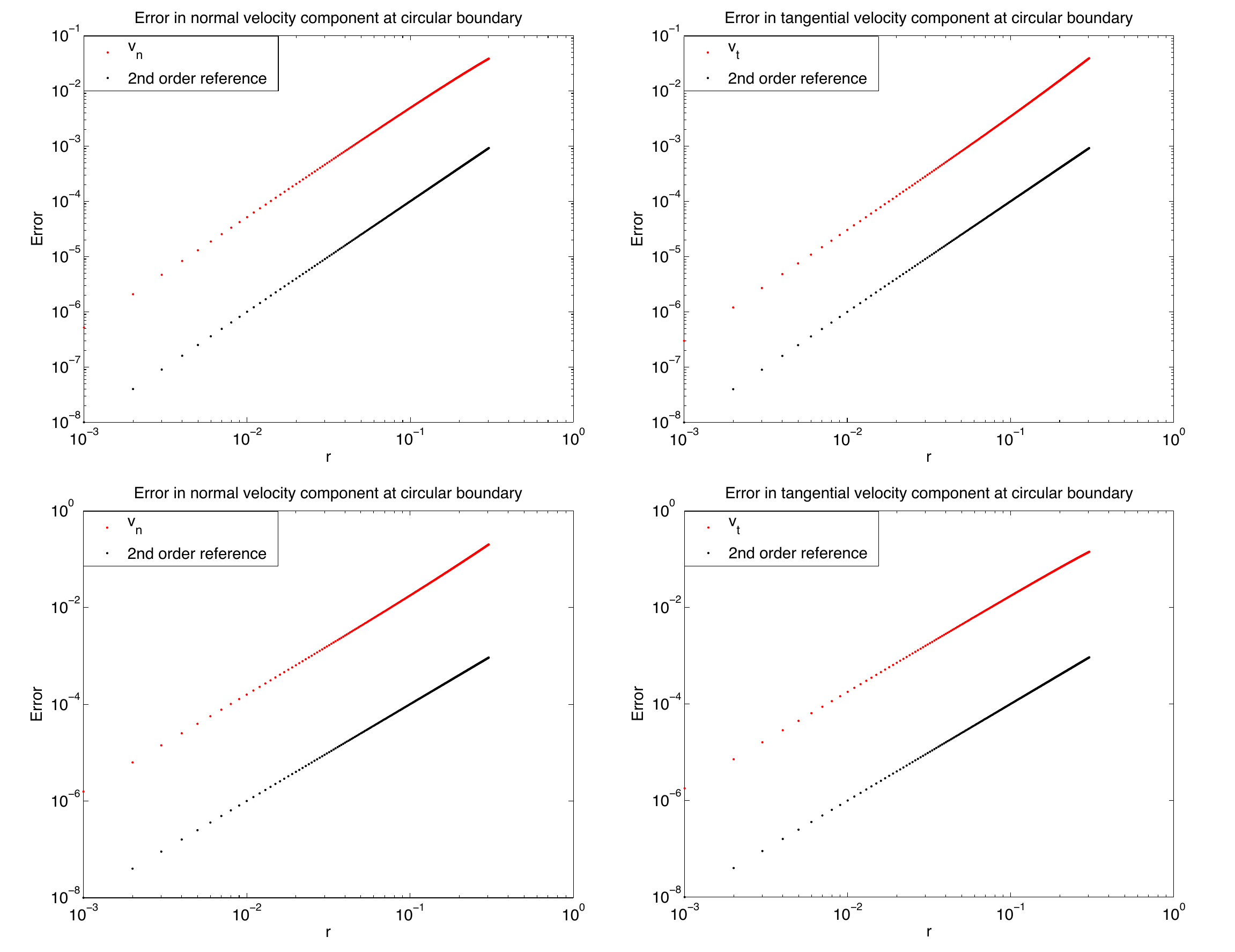}
    \caption{Error in velocity components along circular wall in phase A (upper two figures) and phase B (lower two figures) for the first case}
    \label{fig:boundaryvel1_left}
\end{figure}

\begin{figure}[h!]
  \centering
     \includegraphics[width=0.9\textwidth]{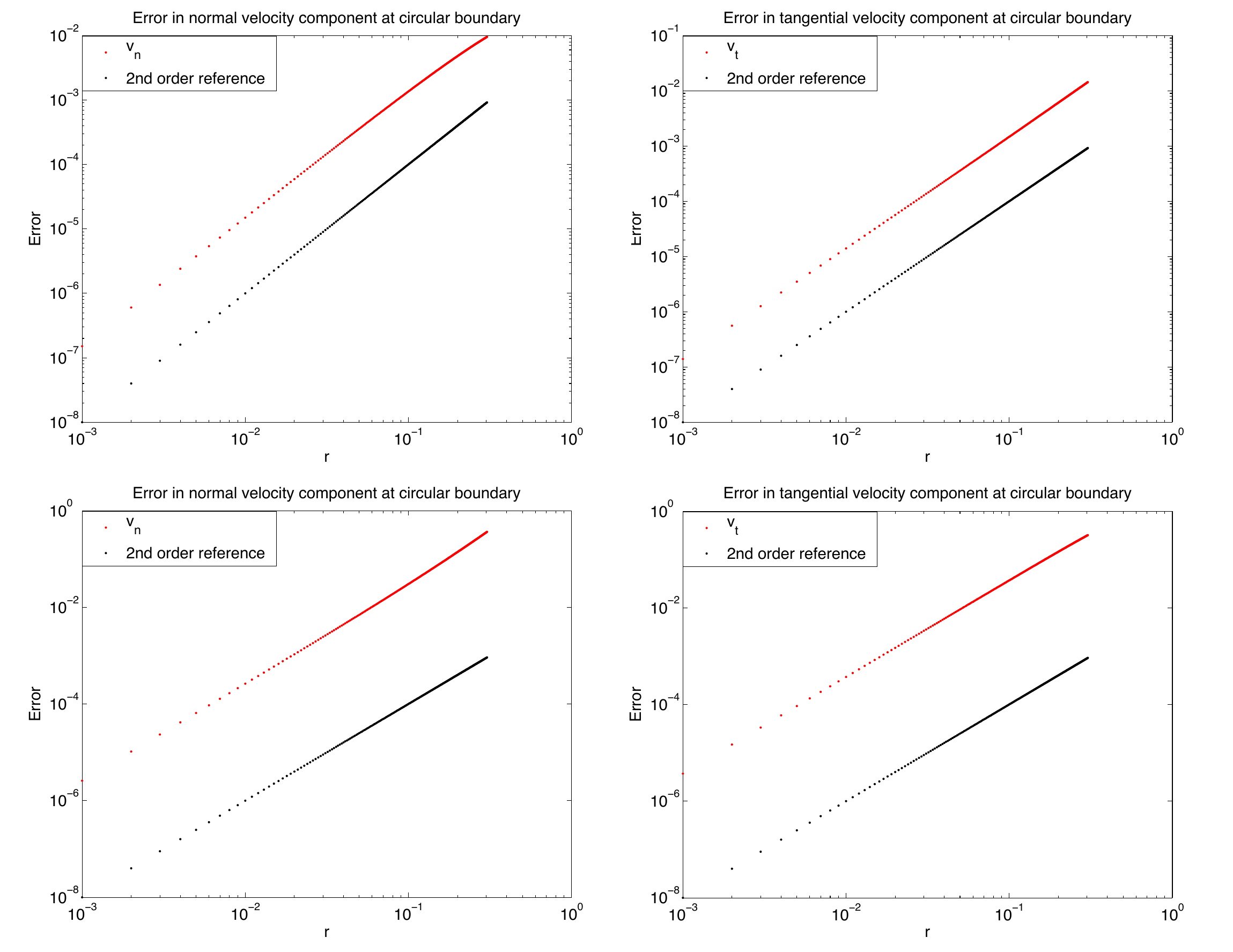}
    \caption{Error in velocity components along circular wall in phase A (upper two figures) and phase B (lower two figures) for the second case}
    \label{fig:boundaryvel2_left}
\end{figure}

\begin{figure}[h!]
  \centering
     \includegraphics[width=0.9\textwidth]{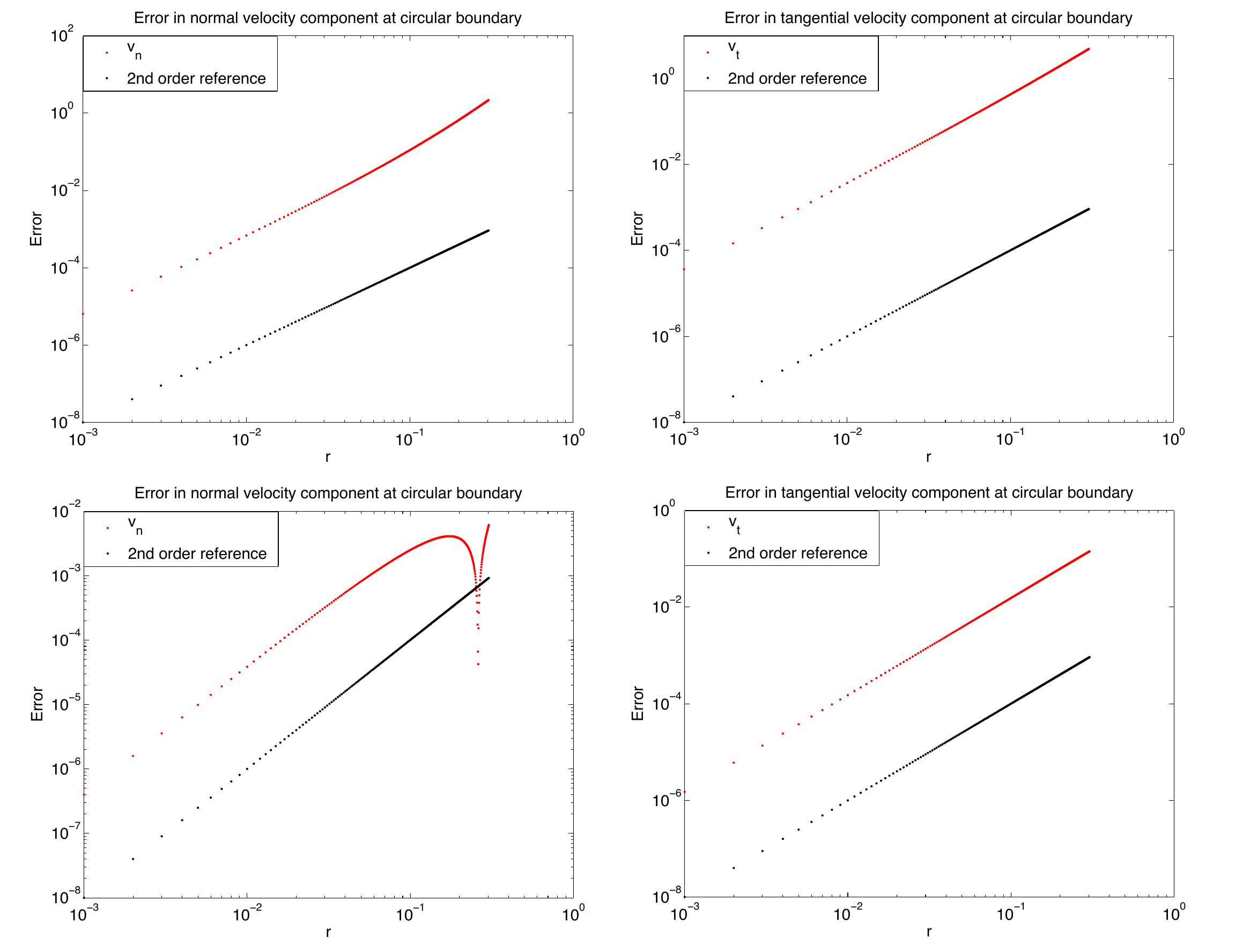}
    \caption{Error in velocity components along circular wall in phase A (upper two figures) and phase B (lower two figures) for the third case}
    \label{fig:boundaryvel3_left}
\end{figure}

\begin{figure}[h!]
  \centering
     \includegraphics[width=0.5\textwidth]{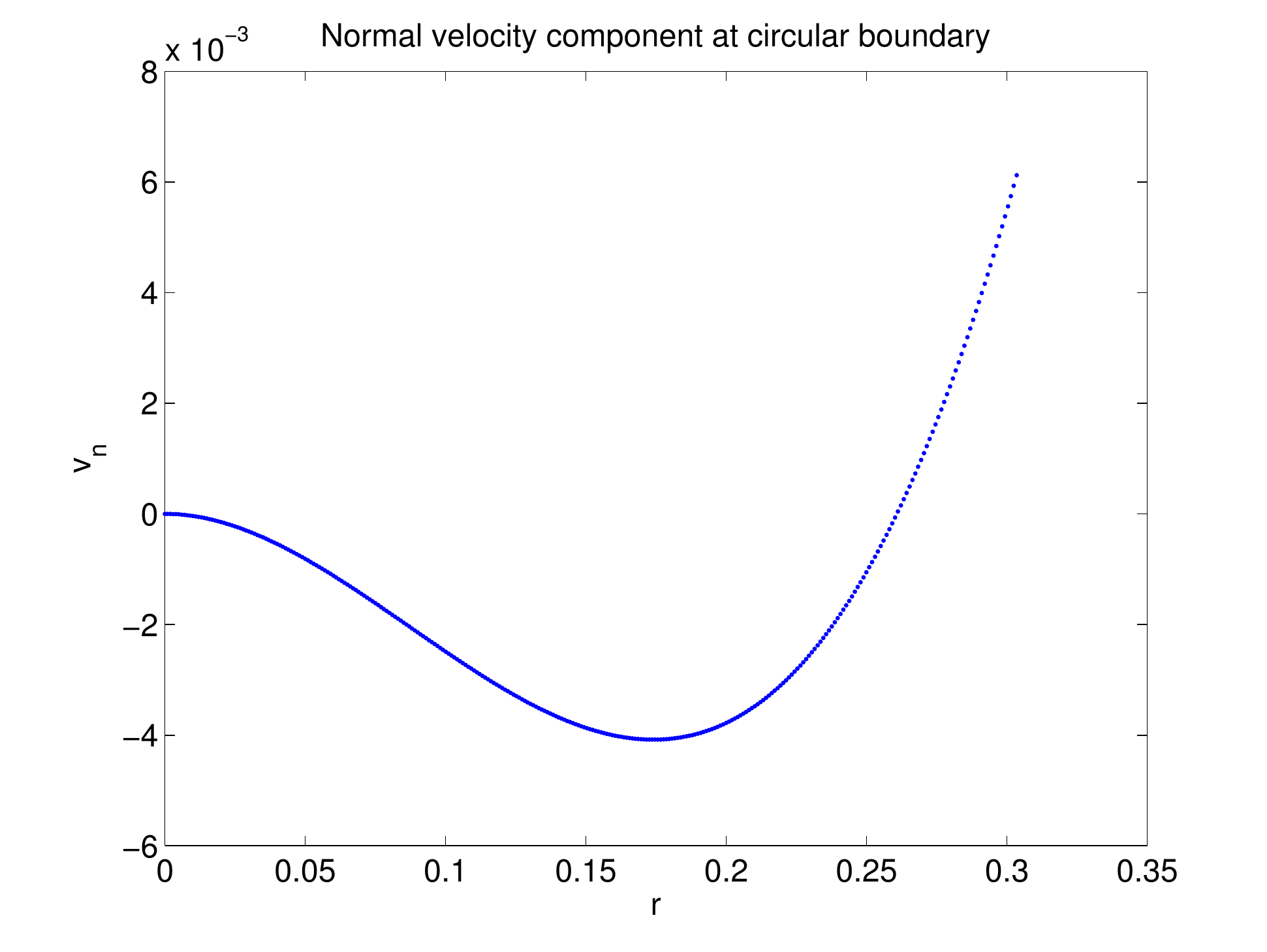}
      \caption{Velocity component tangential to circular wall in phase B for the third case}
    \label{fig:boundaryvel_components}
\end{figure}

To verify the perturbation expansion approximation we consider how well the first two terms satisfy the boundary conditions. The velocity component tangential to the circular wall should be unity, $U=1$, due to the no slip boundary conditions \eqref{eq:BC7} and \eqref{eq:BC8}. Further, the velocity component normal to the wall should be zero according to the boundary conditions \eqref{eq:BC1} and \eqref{eq:BC2}. In \figref{fig:boundaryvel1_left} the error in the normal and tangential velocity components along the circular wall for case one are shown for both fluid phases A and B. The error in the velocity components decrease quadratically with decreasing $r$ ($R$ constant), which agrees with the perturbation expansion approximation (since we have neglected all terms containing powers of $\frac{r}{R}$ of two and higher). Similar second order behavior of the errors in case two and three are shown in \figref{fig:boundaryvel2_left} and \figref{fig:boundaryvel3_left} respectively. However, for case three, i.e. \figref{fig:boundaryvel3_left}, a small deviation of the second order behavior of the error in the velocity component normal to the circular wall in phase B is observed for the larger values of $r$. The reason for the deviation in the second order behavior is that the error changes sign between $r=0.2$ and $r=0.3$. This can bee seen in \figref{fig:boundaryvel_components} where the behavior of the velocity component normal to the circular wall in phase B is plotted. For the smaller values of $r$, approximately $r<0.25$, the normal velocity component is negative, i.e. too small, while for the larger values of $r$ it is too big. For $r> 0.25$ the error again follows a second order behavior. 

The systems for determining the coefficients in the stream function expansion  is singular for $\phi=0$ and $\phi=180$. In case three when $\phi$ is close to $180$ an increase in the error is observed. Compare upper right figures in \figref{fig:boundaryvel2_left} and \figref{fig:boundaryvel3_left} for example. The error in the tangential velocity component along the circular wall for large values of $r$ in phase A is also clearly visible in \figref{fig:vel_pert3}.

For larger values of $r$ than included in the error plots presented here the assumption of a small $\frac{r}{R}$ is no longer valid, and the perturbation expansion approximation fails. Furthermore, far away from the contact point the viscous effects no longer dominate the convection, and the creeping flow approximation also fails. For small values of $r$ on the other hand, the velocity field contains large gradients and is therefore very sensitive to small changes in $r$ or $\theta$.


\newpage
\clearpage
\subsection{Change of Frame of Reference}
The approximated velocity field can be transformed to a frame of reference where the contact point moves over a stationary wall. 
This is done by subtracting a rigid body rotation from the velocity field derived in the previous section. The rigid body rotation is centered at the center of curvature of the solid wall, and given by $r_d\,\Omega\, \fk{e}_\Theta$. Here $\Omega=U/R$ is the magnitude of the angular velocity, $r_d$ is the distance to the  center of curvature of the wall,  and  $\fk{e}_\Theta$ is the  unit vector in the rotational direction. 
The two latter quantities are
\begin{equation}\nonumber
r_d=\sqrt{r^2\cos^2\theta+(R+r\sin\theta)^2},\quad          \fk{e}_\Theta=-\frac{R\cos\theta}{r_d} \, \fk{e}_r      + \frac{R \sin\theta+r}{r_d}  \,  \fk{e}_\theta.
\end{equation}

\begin{figure}[h!]
  \centering
        \includegraphics[width=0.7\textwidth]{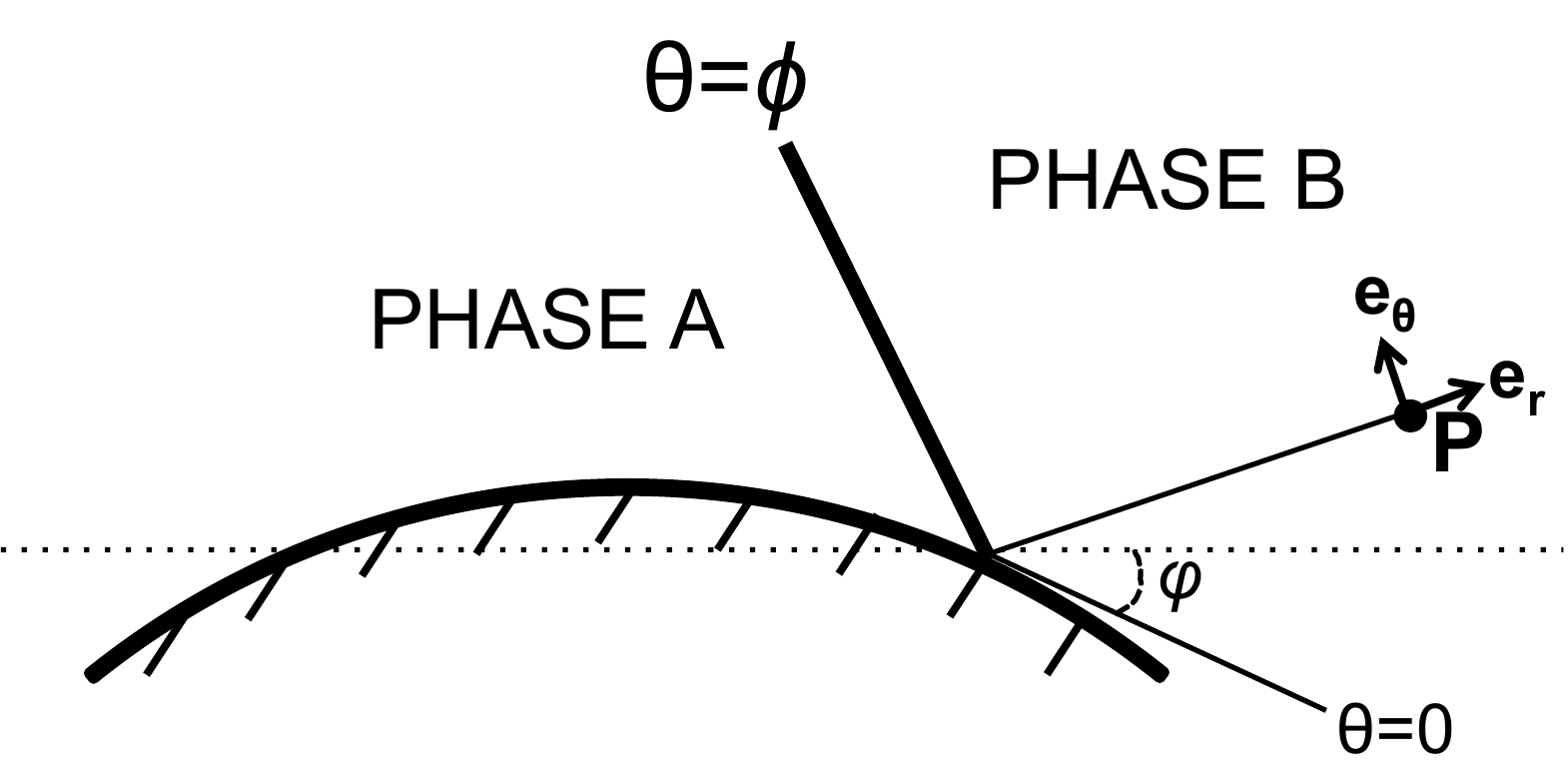}
      \caption{Shifted frame of reference, the contact point is moving over a fixed wall}
    \label{fig:TiltedCoordSyst}
\end{figure}

As demonstrated in \figref{fig:TiltedCoordSyst}, the tangent to the circular wall at the contact point, i.e the line $\theta=0$, will not remain horizontal when the contact point moves over the wall (with the frame of reference of a stationary wall). Therefore we also need to take into account the angle of which the tangent is shifted from a horizontal line when evaluating the polar coordinate $\theta$. The requested $\theta$ is the angle to the horizontal line plus the angle of shift from a horizontal line of the tangent, here denoted by $\varphi$. 

\section{PART \rom{2}: Numerical Simulation of Moving Contact Points}
\label{sec:PART2} 

In this section we demonstrate how the hydrodynamic model derived in previous section could be used in a multiscale model for simulation of moving contact point problems. In the standard model a non-vanishing contact point velocity is in conflict with the standard no-slip boundary condition for the fluid velocity. We will instead impose boundary conditions for the fluid at a small distance from the solid wall. The precise conditions are given by the hydrodynamic solution.

A basic assumption is that there is both a spatial and temporal scale separation between the local microscale contact point behavior and macroscopic fluid flow \cite{MARTIN2}. 
When the dynamics of the moving contact point is driving the flow, the assumption that the flow on the microscale is reacting much faster than the flow on the macroscale is justified \cite{MARTIN2}.
This temporal scale separation implies that the microscopic dynamics is in equilibrium on the time scale of the dynamics of the macroscale. In particular, as the macroscale contact angle evolves, there is an equilibrium microscale contact point speed for each macroscale wall contact angle. Therefore the macroscale contact angle is the only information from the macroscopic model required to determine a contact point velocity. Consequently, there is a relation between the slip velocity of the contact point and the macroscopic wall contact angle and it is this relation that incorporates the microscopic effects at the macroscale.
 
Examples of micro-models that can be used for determining such relations between the contact point velocity and the wall contact angle are molecular dynamics  \cite{REN1,REN2} or phase field models \cite{MARTIN2}.
In order for the multiscale model to be efficient, the microscopic model can for example be used to pre-compute the contact point velocity $U$ for a set of contact angles. The focus here is on how to incorporate the microscopic effects, given by this relation, on the macroscopic dynamics. We will therefore not further discuss the microscopic origin of this relation.

In the proceeding sections we start by discussing the multiscale model in \secref{sec:multiscale} and then present numerical experiments and results in \secref{sec:numexp}.

\subsection{Multiscale Model for the Moving Contact Point}
\label{sec:multiscale}
\begin{figure}[h!]
  \centering
    \includegraphics[width=0.75\textwidth]{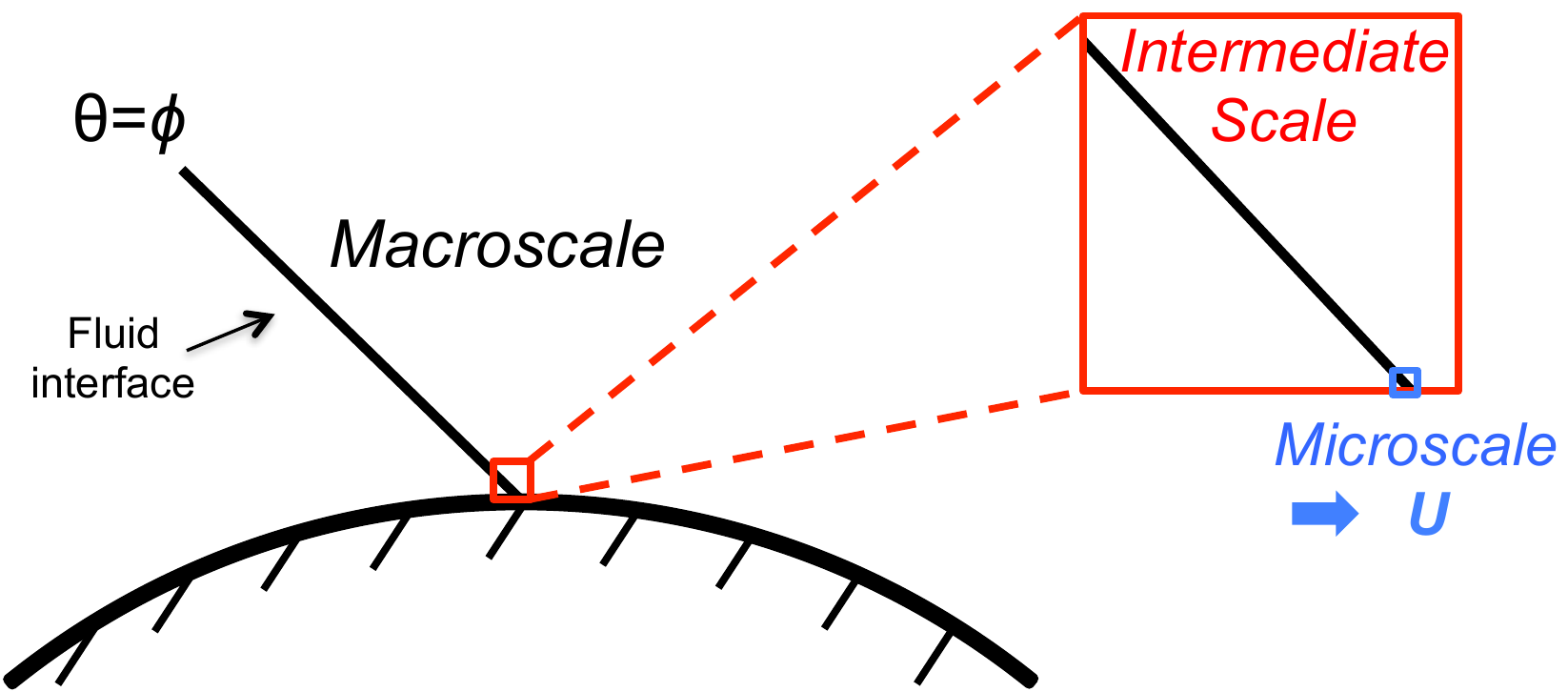}
    \caption{Schematic illustration of the different scales in the multiscale moving contact point model}
    \label{fig:scales}
\end{figure}

A schematic illustration of the different scales of a moving contact point problem is given in \figref{fig:scales}. We introduce an intermediate region of length scale $L$, the red region in \figref{fig:scales}. If a typical length scale of features in the flow is $M$, then $L\ll M$. In this region standard continuum flow is still assumed, but with viscous effects dominating over convective effects, i.e. the Reynolds number $\mathrm{Re} \approx1$. Therefore, the creeping flow approximation of Navier-Stokes equations is valid. Further, at this length scale, the fluid interface is to first approximation flat, which is illustrated in \figref{fig:scales}. Consequently, the hydrodynamic model is a good approximation in this region, and we can use the analytic expressions for the velocity derived in the first part of this paper. The assumptions on which the hydrodynamic model are based are not valid closer to the contact point, see the microscopic view in \figref{fig:scales}. There viscous bending becomes important \cite{EGGERS}, and molecular phenomena such as diffusive mass transport come into play. These microscopic phenomena should be included in the coupled micro-model discussed above. 

We will impose boundary conditions for the fluid velocity at a small distance from the solid wall, in the intermediate region. The precise conditions are given by the hydrodynamic solution.

\subsubsection{Contact Point Boundary Conditions}
\label{sec:CL_BC}
\begin{figure}[h!]
  \centering
    \hspace{0.45cm}\includegraphics[width=0.82\textwidth]{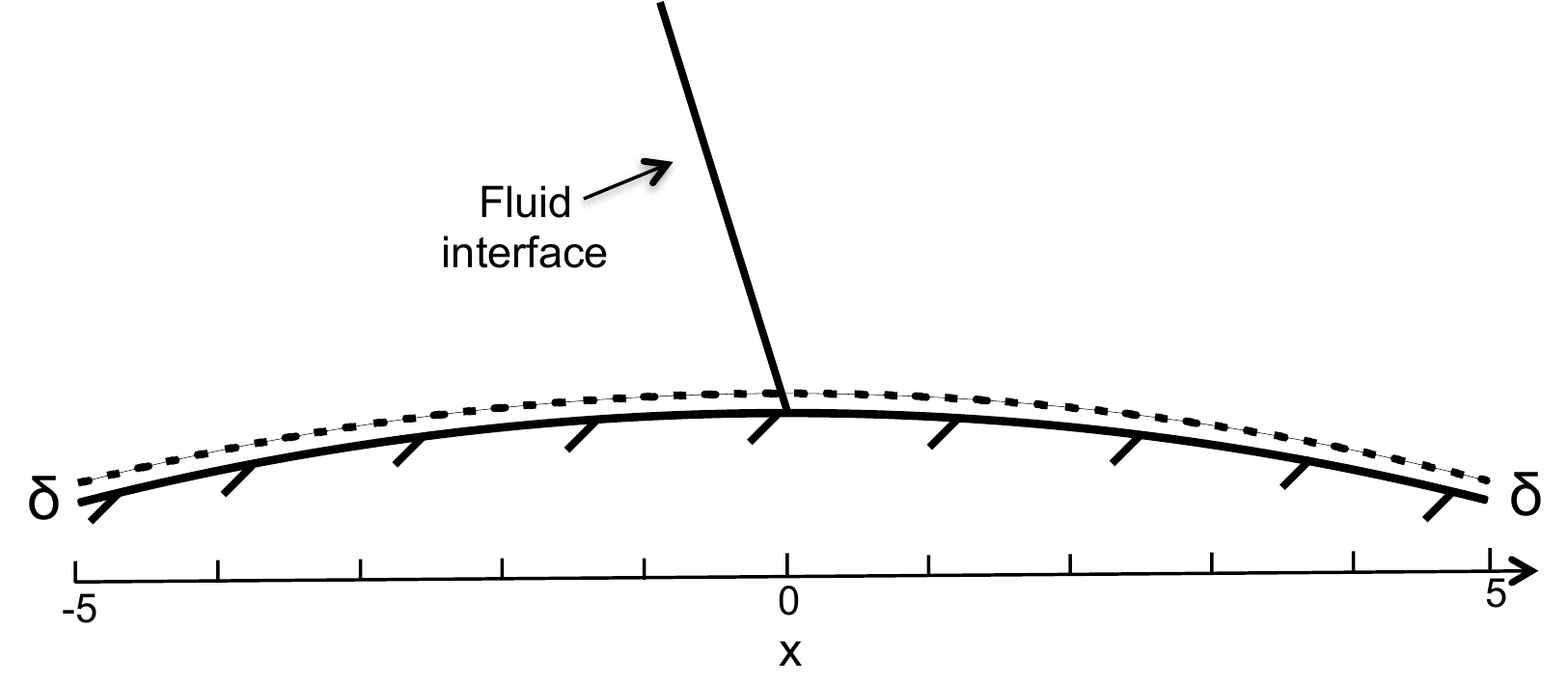}
    \caption{Modified domain. Velocity Dirichlet boundary conditions are applied along the artificial (dashed) boundary}
    \label{fig:mod_dom}
\end{figure}

\begin{figure}[h!]
  \centering
    \includegraphics[width=0.95\textwidth]{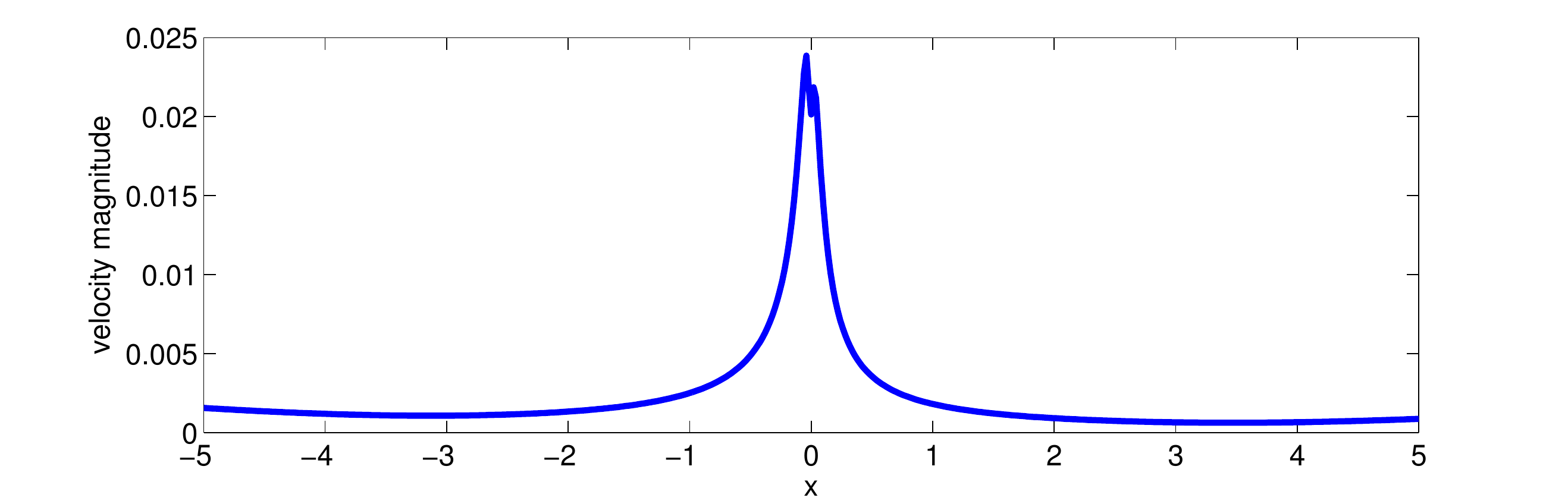}
    \caption{The magnitude of the velocity given by the hydrodynamic model at the artificial boundary}
    \label{fig:fullline}
\end{figure}

The analytical velocity field given by the hydrodynamic model, is used to formulate boundary conditions for the velocity components. Since the hydrodynamic model is not valid exactly at the contact point, the innermost part of the intermediate region will not be included in the simulation. The solution will not be computed in the excluded region. The simplest approach is to use a computational domain that is $\delta$ smaller in the direction perpendicular to the solid wall, see \figref{fig:mod_dom}. Along the new, artificial boundary, which is $\delta$ inside the physical boundary, we impose the analytical velocity from the hydrodynamic model as a Dirichlet boundary condition for the velocity components. The analytical velocity depends on the contact point velocity $U$, which is given by the relation between the macroscopic contact angle and the contact point velocity mentioned above. Thus, the information concerning the movement of one single point (in 2D), is transformed into a velocity boundary condition along the whole boundary. For the domain in \figref{fig:mod_dom} with $\delta= 0.05$, the magnitude of the velocity at the artificial boundary is plotted in \figref{fig:fullline}. When implementing this velocity boundary condition, care has to be taken when choosing the value of $\delta$. The creeping flow approximation is valid in a region with length scale $L$ (the intermediate region), and $ \delta< L$ is required. However, the smaller $\delta$ the sharper is the peak in the boundary velocity function (see \figref{fig:fullline}). Therefore, when discretizing in space the grid must sufficiently resolve the features of the velocity at the boundary, that is $h<\delta$ is required. 


\subsection{Numerical Experiments}
\label{sec:numexp}
The purpose of the numerical experiments is to investigate how well the velocity field based on the boundary conditions from the new hydrodynamic model advects the level set function and hence the contact point. We will consider low Reynolds number 2D flow in a symmetric channel with variable cross section, with an interface separating two identical fluids. We will assume the interface at each moment is in a quasi-steady state shape in the form of a circular arc. Such shapes are formed for instance when liquid rises in a narrow tube due to capillarity.

\subsubsection{Two-Phase Flow Model}
\label{sec:twophase}

The motion of the two immiscible fluids is given by the incompressible Navier--Stokes equations for velocity $\textbf{u}$ and pressure $p$ in non-dimensional form,
\begin{equation}
  \begin{aligned}
  \rho\frac{\partial{\bold u}}{\partial t} + \rho {\bold u} \cdot \nabla {\bold u} &= -
  \nabla p +\frac{1}{\mathrm{Re}}\nabla \cdot (2\mu\nabla^{\mathrm{s}}{\bold u})
  +\mathrm{\textbf{F}}_{st},
  \\
  \nabla \cdot {\bold u}&=0.
\label{eq:nav-stok}
  \end{aligned}
\end{equation}
Here, $\mathrm{\textbf{F}}_{st}$ is the surface tension force at the fluid interface. Further, $\nabla^{\mathrm{s}}\bold{u} = \frac{1}{2}(\nabla \bold u + \nabla \bold u^T)$ denotes the rate of deformation tensor and the parameters $\rho$ and $\mu$ denote the density and viscosity measured relative to the parameters of fluid 1,
\[
\rho = \left\{\begin{array}{ll} 1 & \text{in fluid 1,} \\ \frac{\rho_2}{\rho_1} & \text{in fluid 2,} \end{array} \right. \qquad \mu = \left\{\begin{array}{ll} 1 & \text{in fluid 1,} \\ \frac{\mu_2}{\mu_1} & \text{in fluid 2.} \end{array} \right.
\]

The standard level set method presented in \cite{LEVELSET} is used to keep track of the fluid interface and the moving contact point. The level set function $\phi(\textbf{x}, t)$ is a signed distance function and the fluid interface $\Gamma$ is given by the zero level set of $\phi$, i.e. $\phi = 0$ on the interface. The subdomain $\Omega_1$ occupied by fluid 1 is given by $\phi > 0$ and the subdomain $\Omega_2$ occupied by fluid 2 is given by $\phi < 0$. The level set function is advected in time by the fluid velocity according to the following Hamilton--Jacobi equation
\begin{equation}
\frac{\partial \phi}{\partial t}+  {\bold u} \cdot \nabla \phi= 0.
\label{eq:LS}
\end{equation}
After advecting the fluid interface, the surface tension force $\mathrm{\textbf{F}}_{st}$ is calculated,
\begin{equation}
 \mathrm{\textbf{F}}_{st}= \frac{1}{\mathrm{We}}\kappa  {\bold n}\delta_{{ \Gamma}}, 
\end{equation}
where We is the Weber number and $\delta_{{ \Gamma}}$ is a Dirac delta function with support on $\Gamma$. The normal and curvature of the interface are computed using the level set function,
\begin{align}
{\bold n}=\frac{\nabla \phi}{|\nabla \phi|} , \quad \kappa=-\nabla \cdot {\bold n}. \notag
\label{eq:normcurv}
\end{align}

Over time the level set function will loose its signed distance property due to discretization errors and non-uniform velocity fields. To smooth the level set function and prevent the formation of large gradients, $\phi$ has to be reinitialized with a regular interval.  In section \secref{sec:testproblem} we present our model problem and a reinitialization method specially adapted for this problem. The standard technique for reinitialization (Sussman et.al \cite{REINIT}) consists of solving the following partial differential equation to steady state
\begin{equation}
\frac{\partial \phi}{\partial \tau}=-\mathrm{sign}(\phi_0)(|\nabla \phi|-1),
\label{eq:reinit}
\end{equation}
where $\phi_0$ is the level set function before reinitialization and $\tau$ is a pseudo time step. 
However, solving \eqref{eq:reinit} may result in unphysical volume changes in the fluid phases and does not guarantee preservation of the contact point position. Accurate reinitialization for moving contact lines is a research area itself and is not addressed here.

\subsubsection{Discretization and Implementation}
\label{sec:disc}
We use the two-phase flow solver developed in \cite{MARTINBABA} with suitable modifications to account for moving contact points. The solver is implemented in the C++ based Finite Element open source library deal.ii \cite{DEAL1,DEAL2}. Piecwise continuous linear shape functions on quadrilaterals, $Q_1$ elements, are used for the level set function and Taylor--Hood elements, $Q_2Q_1$ elements, are used for the Navier--Stokes equations (see \cite{MARTINBABA}). For time stepping, each of the level set equation and Navier--Stokes equations are discretized separately using the second order accurate implicit BDF--2 scheme. In order to avoid an expensive coupling between the incompressible Navier--Stokes part and the level set part (via the variables $\textbf{u}$ and $\phi$) a temporal splitting scheme is introduced. 
After time discretization and linearization of the Navier--Stokes equations linear systems need to be solved. For the level set equation, a BiCGStab solver is used due to non-symmetry. The resulting system after discretization of the Navier--Stokes equations is of saddle point structure and solved by an iterative GMRES solver \cite{SAAD}. For preconditioning, a block-triangular operator constructed using the so called Schur complement of the block system is applied from the right \cite{ELMAN}. For more details about the two-phase flow solver we refer to \cite{MARTINBABA}.

\subsubsection{The Test Problem}
\label{sec:testproblem}
\begin{figure}[h!]
  \centering
    \includegraphics[width=0.6\textwidth]{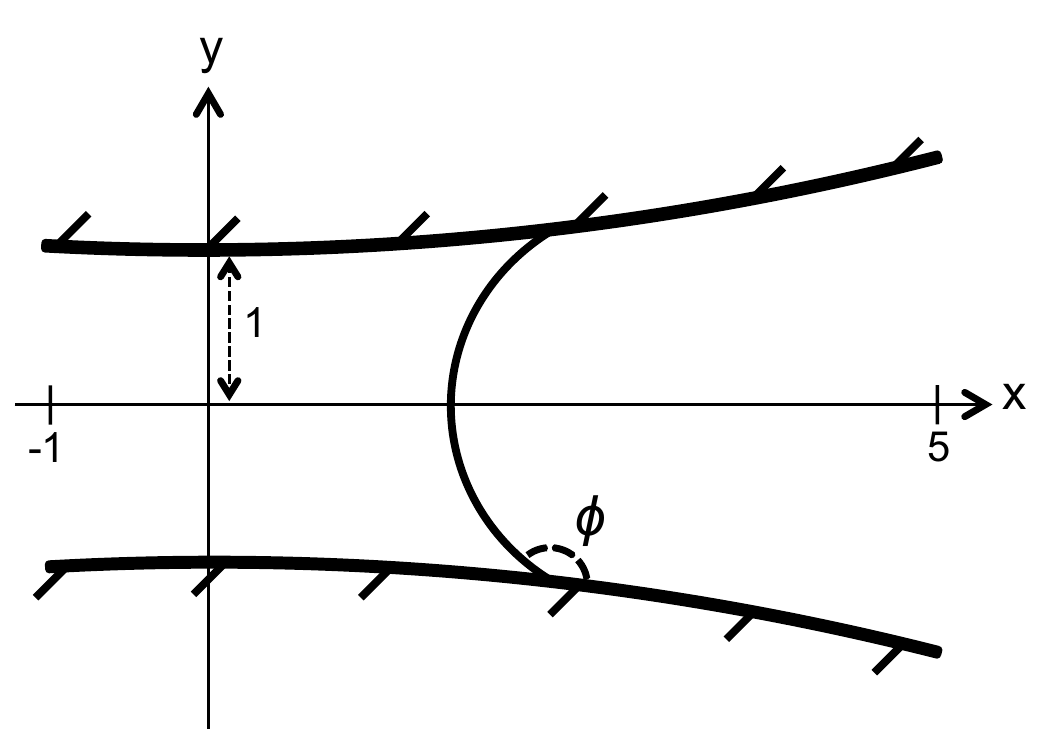}
    \caption{The test problem: a fluid interface is moving in a channel with variable cross section}
    \label{fig:domain}
\end{figure}

The test problem consists of a circular interface moving in a channel with circular walls, see \figref{fig:domain}. The non-dimensional width of the domain is 6, the smallest distance between the walls is 2 and the radius of curvature of the circular walls is 20. Both fluids are assumed to have viscosity $\mu = 0.7$ and density $\rho = 1$, i.e. $Q=1$ in the hydrodynamic model. 
 
The multiscale model does not depend on what microscopic model is used and we do not use a specific relation between the contact angle $\phi$ and velocity $U$, from a specific micro model. Instead we construct hypothetical examples for the relation between $U$ and $\phi$.  

If no reinitialization is used the gradient of the level set function, $\nabla \phi$, gets distorted at the interface close to the contact point.  
We use the following simple reinitialization procedure:
\begin{enumerate}
	\item The contact point position is evaluated by cubic interpolation using the level set function values at the two degrees of freedoms closest to the contact point in each direction along the wall.
	\item The level set function is redefined to represent a distance function to a circular arc interface with the contact point position calculated in previous step.
\end{enumerate}
We emphasize that this reinitialization procedure is adapted to our idealized test problem and is not useful in most other, more realistic settings.

\subsubsection{Results}
\begin{figure}[h!]
  \centering
    \includegraphics[width=0.8\textwidth]{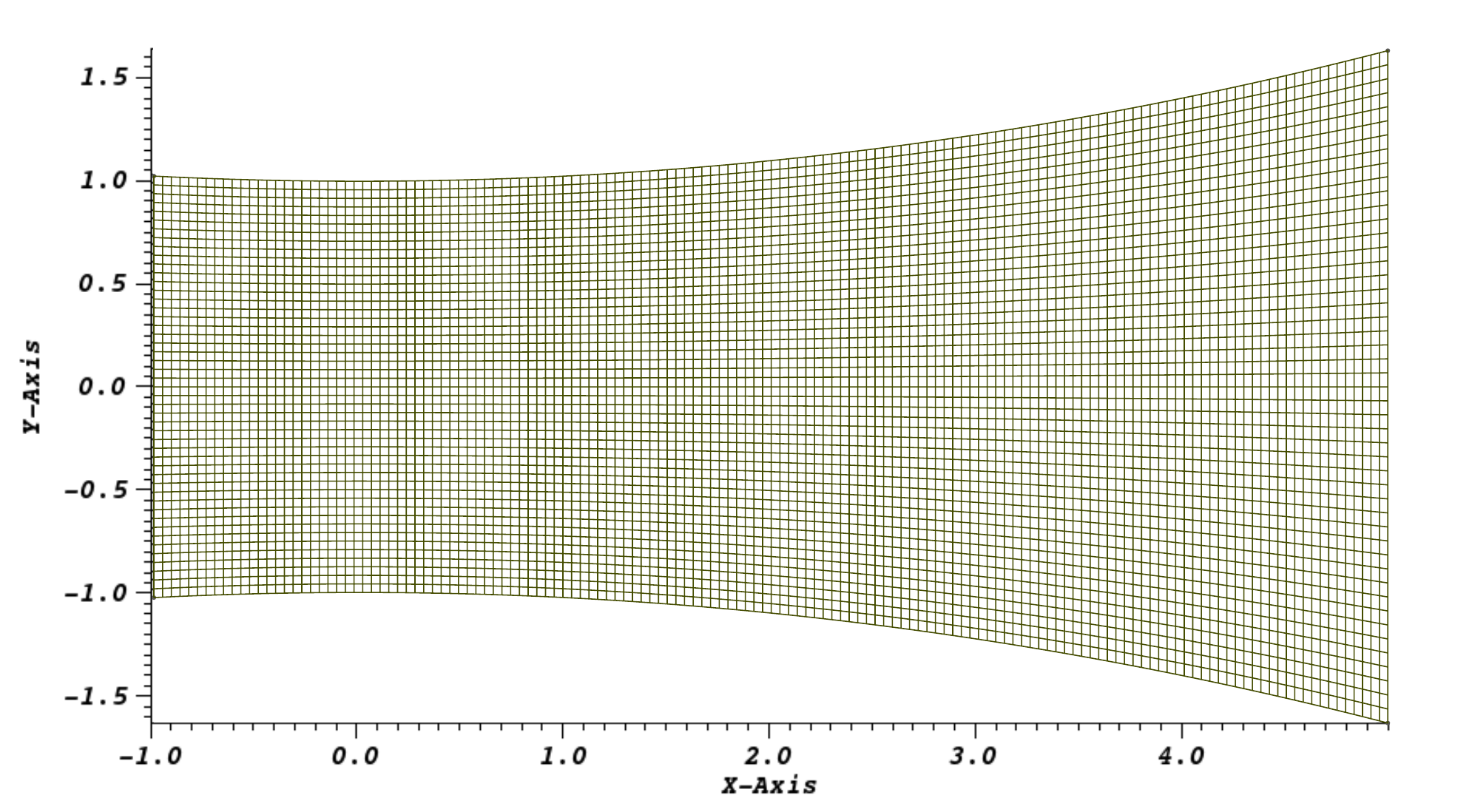}
    \caption{\figtext{Structured mesh with smallest element length $h=1/24$.}}
    \label{fig:grid}
\end{figure}

We perform two sets of simulations. In the first set the wall contact angle is $\phi = 140$  and the initial contact point position is $x_{initial}=3$. In the second set $\phi = 170$ and $x_{initial}=0$. We hypothetically relate both contact angles to a contact point velocity of $U = 0.02$. We investigate spatial grid convergence by performing simulations using a set of structured meshes with different sizes. The smallest element length for each mesh is $h = 1/24,1/32, 1/40, 1/48$ and $1/56$ respectively. The mesh where $h=1/24$ is depicted in \figref{fig:grid}. For $\phi = 170$ the coarsest mesh $h=1/24$ is not used.  

\begin{figure}[h!]
  \centering
    \includegraphics[width=0.7\textwidth]{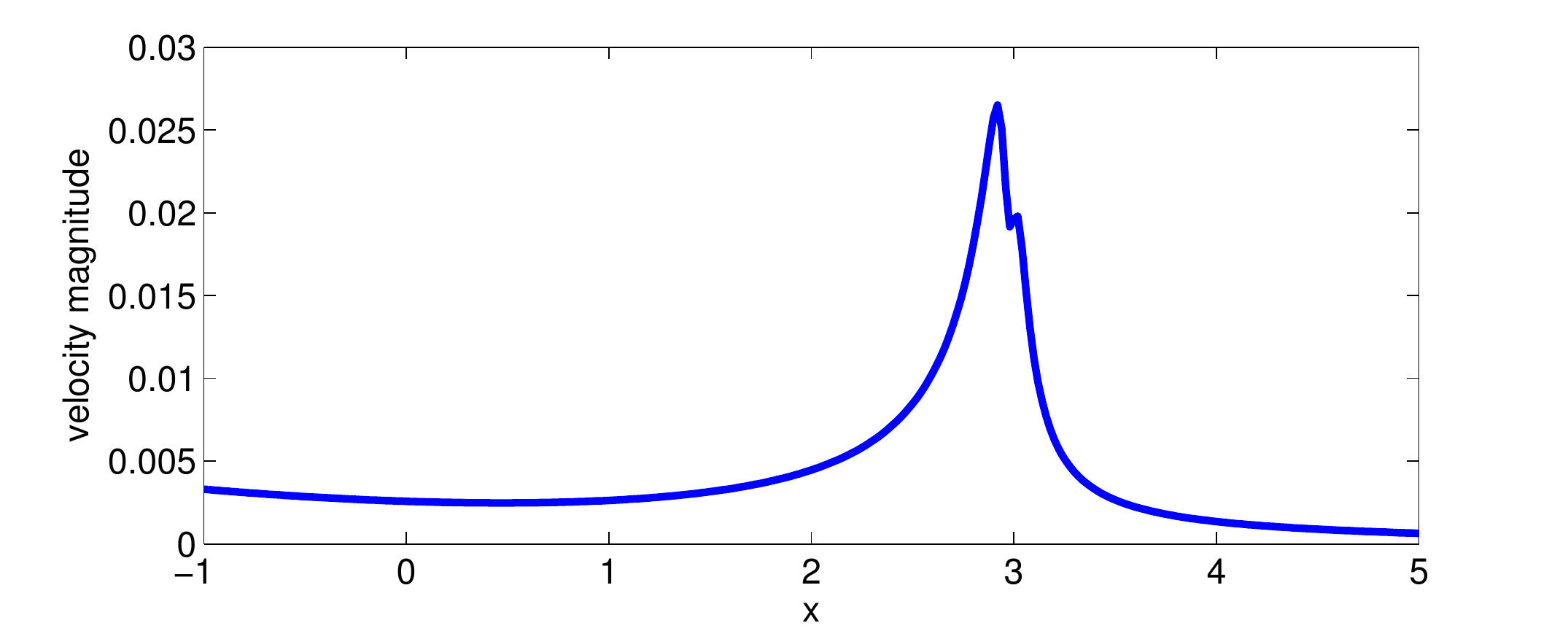}
    \caption{Magnitude of the velocity boundary function applied at the artificial boundary ($\delta = 0.05$) when $\phi=140$}
    \label{fig:BC_FN140}
\end{figure}

\begin{figure}[h!]
  \centering
    \includegraphics[width=0.7\textwidth]{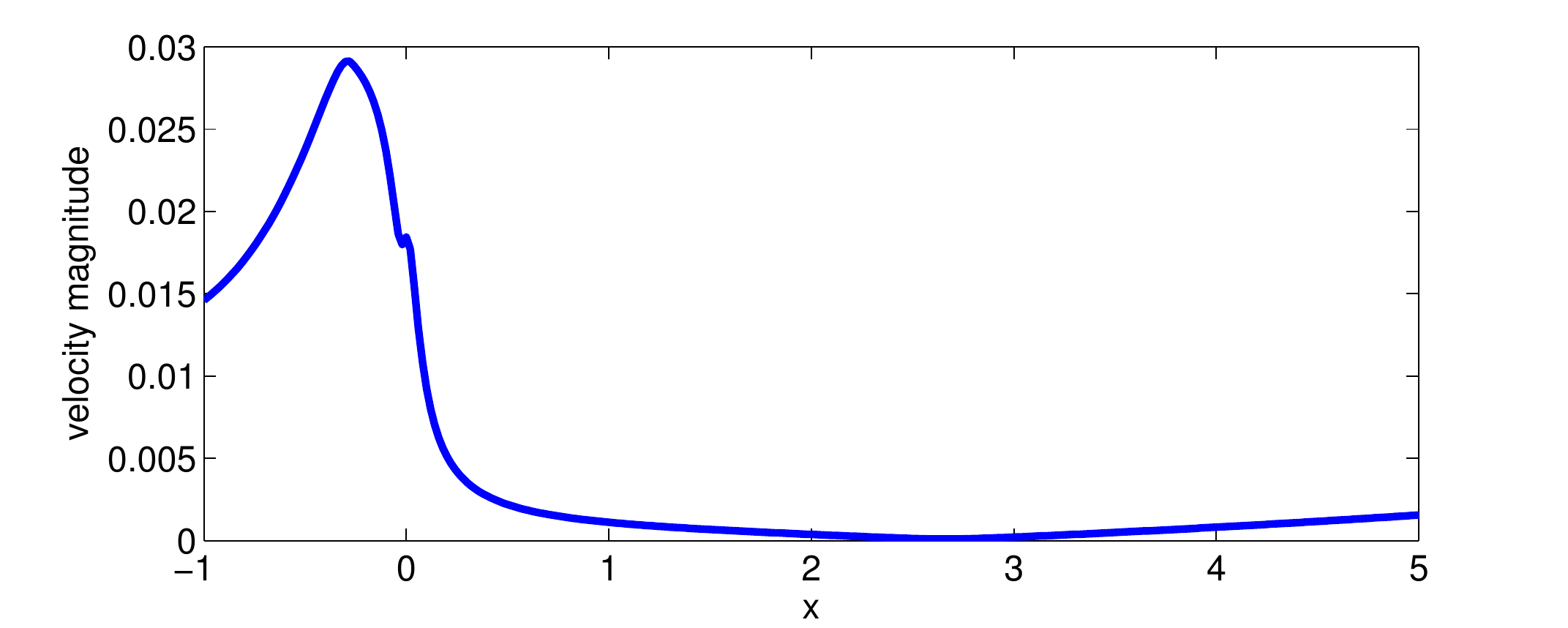}
    \caption{Magnitude of the velocity boundary function applied the artificial boundary ($\delta = 0.05$) when $\phi=170$}
    \label{fig:BC_FN170}
\end{figure}

As explained in \secref{sec:CL_BC}, when choosing $\delta$ (the distance between the artificial and physical boundary) we need to make sure $\delta < L$, where
$L$ is the characteristic length scale of the intermediate region. For the creeping flow to be
valid we need $\mathrm{Re} = \frac{\rho UL}{\mu} \ll 1$ or $L \ll \frac{\mu}{\rho U}$. For this model problem this implies
the following condition on the distance to the physical boundary: $\delta < L \ll 35$, and we use $\delta = 0.05$. The velocity functions that are used as boundary conditions at the artificial boundaries are shown in \figref{fig:BC_FN140} and \figref{fig:BC_FN170} for $\phi = 140$ and $\phi = 170$ respectively. We use homogenous Neumann boundary conditions for the velocity at the left and right boundaries. 


\begin{figure}[h!]
  \centering
    \includegraphics[width=0.7\textwidth]{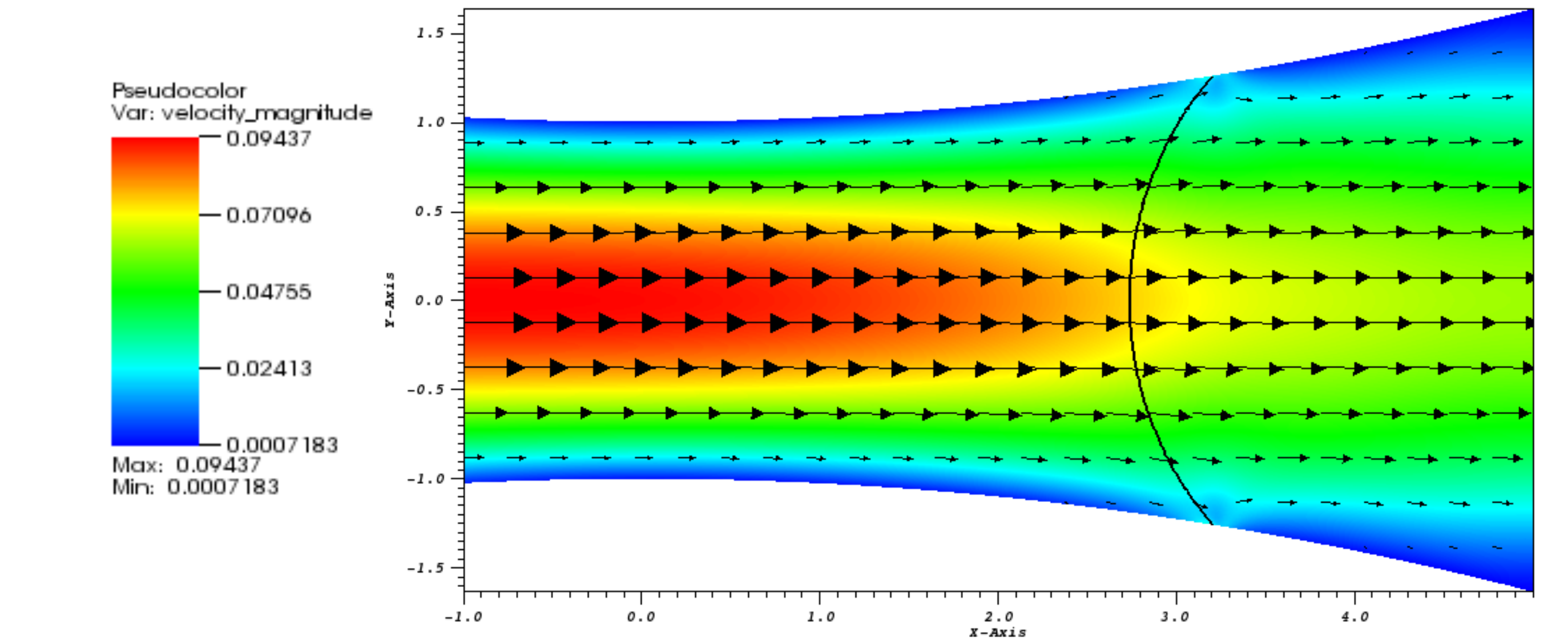} 
    \caption{Resulting velocity fields at T=10. $\phi = 140$, $U =0.02$, $Q=1$}
    \label{fig:ResultPlots}
\end{figure}

\begin{figure}[h!]
  \centering
    \includegraphics[width=0.7\textwidth]{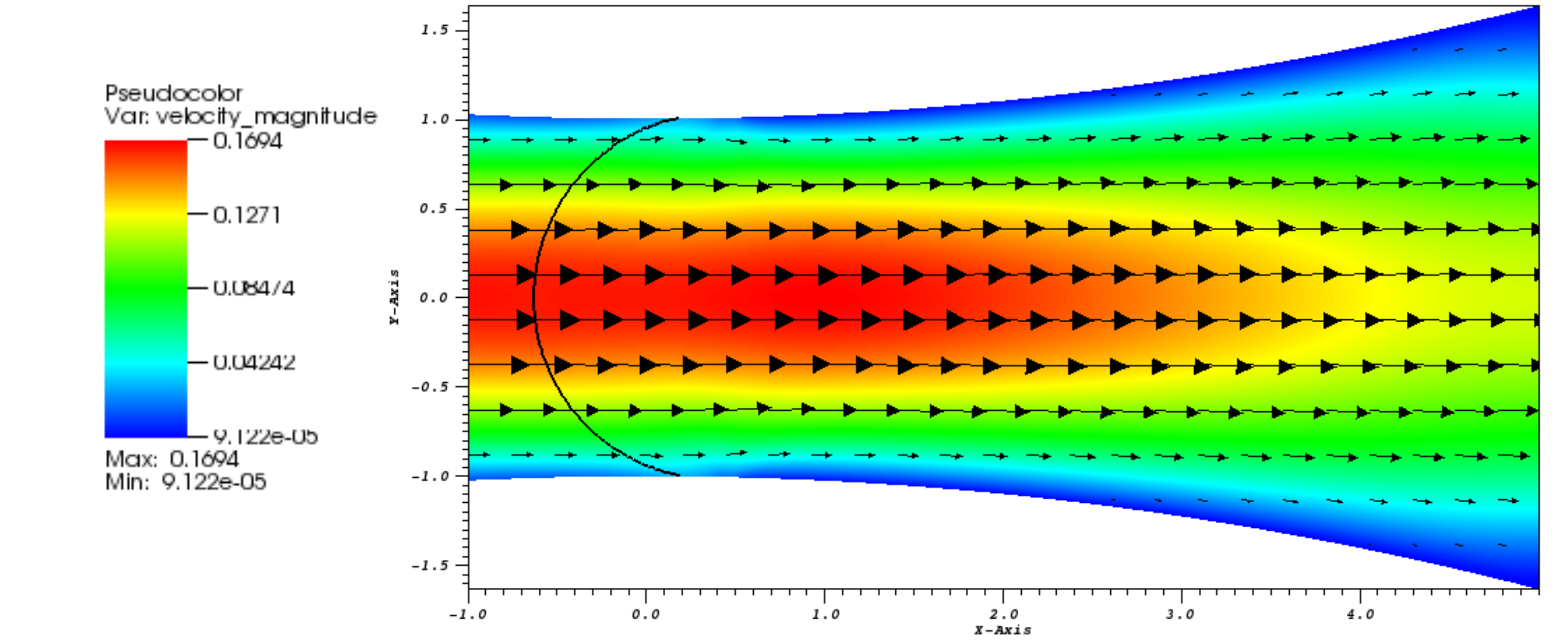}
    \caption{Resulting velocity fields at T=10. $\phi = 170$, $U =0.02$, $Q=1$}
    \label{fig:ResultPlots2}
\end{figure}

The simulations are run for a non-dimensional time of $T=10$ with a time step size of $\Delta t = 0.0001$ for the case when $\phi=140$ and $\Delta t = 0.005$ when $\phi=170$. We are interested in the spatial errors when investigating the velocity boundary conditions and with this time step sizes the spatial discretization errors are assumed to dominate the temporal. The resulting velocity fields at $T$ for the finest mesh ($h = 1/56$) are plotted in \figref{fig:ResultPlots} and \figref{fig:ResultPlots2}. It can be seen that away from the interface the flow profile is a regular poiseuille profile with velocity very close to zero at the artificial boundary. Close to the interface and the contact point however, the velocity is far from zero at the artificial boundary due to the velocity boundary condition. 


\begin{figure}[h!]
  \centering
    \includegraphics[width=1\textwidth]{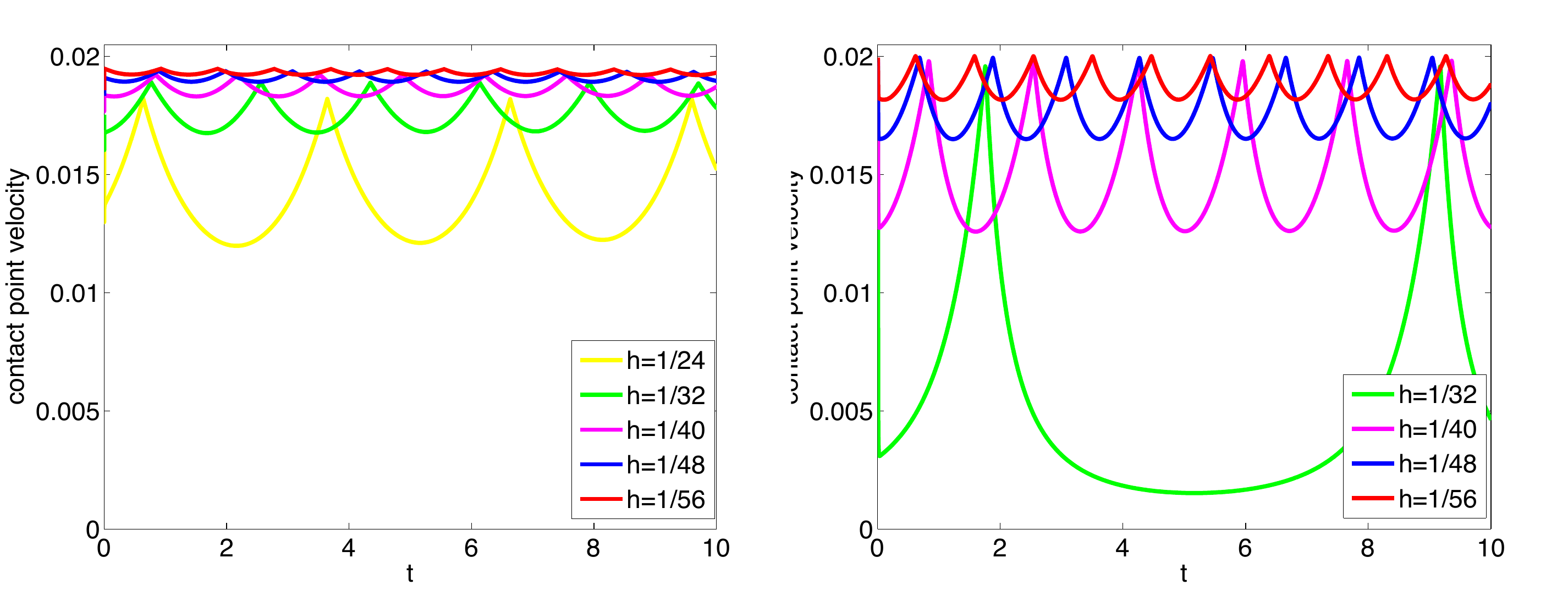}
    \caption{Resulting contact point velocities, $\phi = 140$ (left) and $\phi = 170$ (right)}
    \label{fig:velocities}
\end{figure}

The resulting contact point velocity in each time step is plotted as a function of time in \figref{fig:velocities}. The period of the oscillations corresponds to the time it takes for the contact point to pass one grid cell, except for the coarsest mesh when $\phi = 170$ (i.e. $h=1/32$). For this case it seems that the mesh is not capable of accurately resolving the velocity boundary function. The error in the velocity is higher for the case with the larger contact angle $\phi=170$. This agrees with the results presented in \secref{sec:result1} where it can be seen that the error in the velocity field from the hydrodynamic model is higher for the case of a larger contact angle. 

\begin{figure}[h!]
  \centering
    \includegraphics[width=1\textwidth]{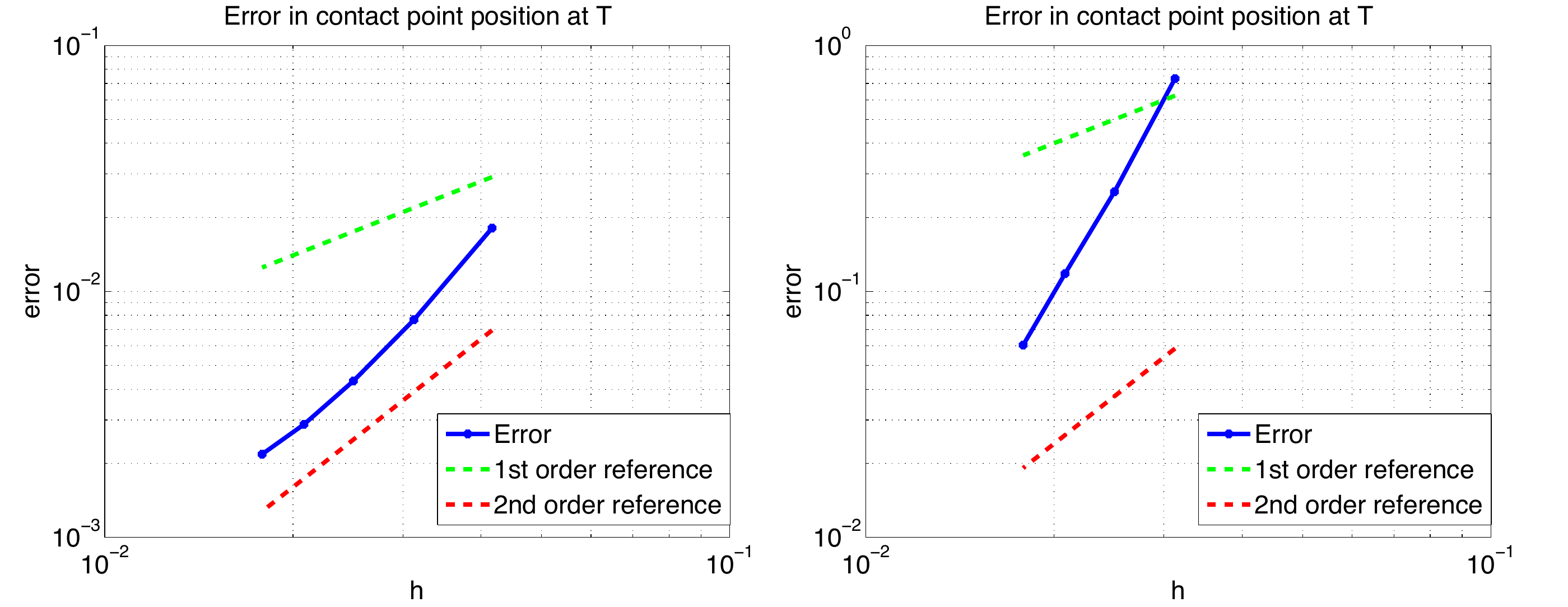}
    \caption{Error in contact point position at T = 10, $\phi = 140$ (left) and $\phi = 170$ (right)}
    \label{fig:conv}
\end{figure}

We measure the relative error in the contact point position at time $T$ compared to the correct position $UT$ for the different grid sizes. In \figref{fig:conv} it can be seen that grid convergence is obtained with a rate of convergence of at least $2$ for both contact angles.
 

\section{Summary and Conclusions}

We have derived a two dimensional hydrodynamic model for the fluid flow near a contact point moving along a curved boundary with local radius of curvature $R$ at the contact point. The model is valid at length scales $L$ and by assuming $R>L$ we have neglected centrifugal and Coriolis forces. 
The model is based on the Stokes equations and was derived using perturbation analysis. As for the original flat wall hydrodynamic model \cite{HUH}, the model is not valid in the immediate vicinity of the contact point, at lengths scales where molecular phenomena come into play.  

We have also discussed a new idea for imposing non-singular Dirichlet boundary conditions for the velocity near a moving contact point (in a two-phase flow simulation). Numerical experiments are presented for a two dimensional model problem of two-phase flow with a moving contact point. Here the non-singular velocity boundary conditions are used together with a level set description of the interface. The results from these simulations demonstrate that the velocity field based on the non-singular boundary conditions is capable of accurately advecting the contact point.

The numerical simulations are based on the idea of excluding a small area around the contact point, and impose the non-singular Dirichlet boundary conditions at an artificial boundary. In this paper we have demonstrated and investigated the idea by excluding a thin region along the whole boundary. However, in a more realistic setting the artificial boundary should only exist along a $\mathcal{O}(L)$ part of the physical boundary. Further, we have applied a special reinitialization, which would not be applicable in most realistic cases. Investigating possible reinitilization strategies is a future research topic. Finally, it is important to note that the model for the moving contact point can also be combined with an interface tracking method as well.

\section*{Appendix}
The equations for the coefficients in the first term in the stream function expansion (\ref{eq:psi}) are written in matrix form as
\begin{equation}
M^0Z^0=F^0
\label{eq:zero_syst}
\end{equation}
where
\begin{equation}
\tiny
M^0= 
\begin{pmatrix}
 0 & 1 & 0 & \pi & 0 & 0 & 0 & 0  \\  
 0 & 0 & 0 & 0 & 0 & 1 & 0 & 0  \\    
 S & C & \phi S & \phi C & 0 & 0 & 0 & 0  \\  
 0 & 0 & 0 & 0 & S & C & \phi S & \phi C  \\ 
 -C & S & -(S+\phi C) & -(C-\phi S) & C & -S & (S+\phi C) & (C-\phi S)  \\ 
 -\mu_AS & -\mu_AC & \mu_A(2C-\phi S) & -\mu_A(2S+\phi C) & \mu_BS & \mu_BC & -\mu_B(2C-\phi S) &\mu_B(2S+\phi C)  \\  
  1 & 0 & \pi & 1 & 0 & 0 & 0 & 0 \\   
 0 & 0 & 0 & 0 & -1 & 0 & 0 & -1    
 \end{pmatrix} \notag, 
 \end{equation}
 
 \begin{equation}
  \scriptsize
 Z^0=
  \begin{pmatrix}		      
    a_A\\
    b_A  \\
    c_A  \\
    d_A  \\
    a_B \\
    b_B \\
    c_B \\
    d_B
 \end{pmatrix}
~~~\text{       and       } ~~~
F^0=
 \begin{pmatrix}		      
    0\\
    0  \\
    0  \\
    0  \\
    0 \\
    0 \\
    -U \\
    U
 \end{pmatrix}.
 \notag
\end{equation}
and 
\begin{align}
&S=\sin\phi \notag \\
&C=\cos\phi .\notag 
\end{align}
The rows in system \eqref{eq:zero_syst} originates from the boundary conditions \eqref{eq:BC1}-\eqref{eq:BC8} respectively. 
The solution is 
 \begin{align}
 &a_A=-U-\pi c_A-d_A  \\
 &b_A=-\pi d_A \\
 &c_A=US^2[S^2-\delta\phi+Q(\phi^2-S^2)]/D   \\
 &d_A=USC[S^2-\delta\phi+Q(\phi^2-S^2)-\pi \tan\phi]/D \\
 &a_B=-U-d_B  \\
 &b_B=0 \\
 &c_B=US^2[S^2-\delta^2+Q(\delta\phi-S^2)]/D  \\
 &d_B=USC[S^2-\delta^2+Q(\delta\phi-S^2)-Q\pi \tan \phi]/D  ,
 \end{align}
where 
\begin{align}
&\delta=\phi-\pi \notag \\
&Q=\mu_A/\mu_B \notag \\
&D=(SC-\phi)(\delta^2-S^2)+Q(\delta-SC)(\phi^2-S^2) .\notag 
\end{align}
Note that the system is non-singular for $0<\phi<180$.


\bibliographystyle{spmpsci}      
\bibliography{references}   

%
%

\end{document}